\documentclass[longauth]{aa}
\usepackage{txfonts}
\usepackage[dvips]{graphics} 
\usepackage{graphicx}
\usepackage{longtable}
\usepackage{color}
\usepackage{dcolumn} 

\newcommand{\pks}{PKS\,2155$-$304}

\newcommand{\percmsqrsT}{\ensuremath{\mathrm{cm}^{-2}\mathrm{s}^{-1}\mathrm{TeV}^{-1}}}
\newcommand{\percmsqrs}{\ensuremath{\mathrm{cm}^{-2}\mathrm{s}^{-1}}}

\begin{document}

\title{VHE $\gamma$-ray emission of PKS\,2155$-$304:\\spectral and temporal variability}

\author{ HESS Collaboration
 \and A.~Abramowski \inst{4}
 \and F.~Acero \inst{15}
 \and F. Aharonian\inst{1,13}
 \and A.G.~Akhperjanian \inst{2}
 \and G.~Anton \inst{16}
 \and U.~Barres de Almeida \inst{8} \thanks{supported by CAPES Foundation, Ministry of Education of Brazil}
 \and A.R.~Bazer-Bachi \inst{3}
 \and Y.~Becherini \inst{12}
 \and B.~Behera \inst{14}
 \and W.~Benbow \inst{1}
% ...
 \newpage
}

\institute{
Max-Planck-Institut f\"ur Kernphysik, P.O. Box 103980, D 69029
Heidelberg, Germany
\and
 Yerevan Physics Institute, 2 Alikhanian Brothers St., 375036 Yerevan,
Armenia
\and
Centre d'Etude Spatiale des Rayonnements, CNRS/UPS, 9 av. du Colonel Roche, BP
4346, F-31029 Toulouse Cedex 4, France
\and
Universit\"at Hamburg, Institut f\"ur Experimentalphysik, Luruper Chaussee
149, D 22761 Hamburg, Germany
\and
Institut f\"ur Physik, Humboldt-Universit\"at zu Berlin, Newtonstr. 15,
D 12489 Berlin, Germany
\and
LUTH, Observatoire de Paris, CNRS, Universit\'e Paris Diderot, 5 Place Jules Janssen, 92190 Meudon, 
France
\and
CEA Saclay, DSM/IRFU, F-91191 Gif-Sur-Yvette Cedex, France
\and
University of Durham, Department of Physics, South Road, Durham DH1 3LE,
U.K.
\and
Unit for Space Physics, North-West University, Potchefstroom 2520,
    South Africa
\and
Laboratoire Leprince-Ringuet, Ecole Polytechnique, CNRS/IN2P3,
 F-91128 Palaiseau, France
\and 
Laboratoire d'Annecy-le-Vieux de Physique des Particules,
Universit\'{e} de Savoie, CNRS/IN2P3, F-74941 Annecy-le-Vieux,
France
\and
Astroparticule et Cosmologie (APC), CNRS, Universit\'{e} Paris 7 Denis Diderot,
10, rue Alice Domon et L\'{e}onie Duquet, F-75205 Paris Cedex 13, France
\thanks{UMR 7164 (CNRS, Universit\'e Paris VII, CEA, Observatoire de Paris)}
\and
Dublin Institute for Advanced Studies, 5 Merrion Square, Dublin 2,
Ireland
\and
Landessternwarte, Universit\"at Heidelberg, K\"onigstuhl, D 69117 Heidelberg, Germany
\and
Laboratoire de Physique Th\'eorique et Astroparticules, 
Universit\'e Montpellier 2, CNRS/IN2P3, CC 70, Place Eug\`ene Bataillon, F-34095
Montpellier Cedex 5, France
\and
Universit\"at Erlangen-N\"urnberg, Physikalisches Institut, Erwin-Rommel-Str. 1,
D 91058 Erlangen, Germany
% ...
}

\abstract  
{Observations of very high energy $\gamma$-rays from blazars 
  provide information about acceleration mechanisms occurring in
  their innermost regions. Studies of variability in these objects
  allow a better understanding of the mechanisms at play.  }
{To investigate the spectral and temporal variability of VHE ($>100\,\mathrm{GeV}$) $\gamma$-rays 
of the well-known high-frequency-peaked BL\,Lac object \pks\ with the H.E.S.S. imaging atmospheric 
Cherenkov telescopes over a wide range of flux states.}
{Data collected from 2005 to 2007 are analyzed. Spectra are derived on
  time scales ranging from 3 years to 4 minutes. Light curve
  variability is studied through doubling timescales and structure
functions, and is compared  with red noise process simulations.}
{The source is found to be in a low state from 2005 to 2007, except for a set of exceptional flares 
which occurred in July 2006. The quiescent state of the source is characterized by an associated mean flux 
level of $(4.32\pm0.09_\mathrm{stat}\pm0.86_\mathrm{syst}) \times 10^{-11}\,{\rm cm^{-2}}\,{\rm s^{-1}}$ 
above $200\,{\rm GeV}$, or approximately $15\%$ of the Crab Nebula, and a power law photon 
index of $\Gamma=3.53\pm0.06_\mathrm{stat}\pm0.10_\mathrm{syst}$. During the flares 
of July 2006, doubling timescales of $\sim 2\,{\rm min}$ are found. The spectral index variation is 
examined over two orders of magnitude in flux, yielding different behaviour at low and high fluxes,
which is a new phenomenon in VHE $\gamma$-ray emitting blazars.
The variability amplitude characterized by the fractional r.m.s. $F_{\rm var}$  is strongly energy-dependent 
and is $\propto E^{0.19\pm0.01}$.
The light curve r.m.s. correlates with the flux. This is the signature of a multiplicative process which can be
accounted for as a red noise with a Fourier index of $\sim 2$.}
{This unique data set shows evidence for a low level
  $\gamma$-ray emission state from \pks, which possibly has a different origin
  than the outbursts. The discovery of the light curve lognormal
  behaviour might be an indicator of the origin of
  aperiodic variability in blazars.}

\offprints{santiago.pita@apc.univ-paris7.fr and francesca.volpe@mpi-hd.mpg.de/volpe@llr.in2p3.fr}

\maketitle

\section{Introduction} \label{intro}

The BL Lacertae (BL Lac) category of Active Galactic Nuclei (AGN) represents the
vast majority of the population of energetic and extremely variable
extragalactic very high energy $\gamma$-ray emitters. Their luminosity varies in
unpredictable, highly irregular ways, by orders of magnitude, and at all
wavelengths across the electromagnetic spectrum. The very high energy (VHE,
$E\geq100\,{\rm GeV}$) $\gamma$-ray fluxes vary often on the shortest timescales
that can be seen in this type of object, with large amplitudes
which can dominate the overall output. It hence indicates that the understanding
of this energy domain is the most important one for understanding the underlying fundamental
variability and emission mechanisms at play in high flux states. 

It has been, however, difficult to ascertain whether $\gamma$-ray emission is present only 
during high flux states or  also when the source is in a more stable or quiescent state  
but with a flux which is  below the instrumental limits.
The advent of the current generation of atmospheric
Cherenkov telescopes with unprecedented sensitivity in the VHE regime gives
new insights into these questions. 

The high frequency peaked BL Lac object (HBL) \pks, located at a redshift $z=0.117$, initially
discovered as a VHE $\gamma$-ray emitter by the Mark 6 telescope (\cite{cha99}), has been detected by the
first H.E.S.S. telescope in 2002-2003 (\cite{HESS2155_2003}~2005b).  It has been frequently observed by the full array
of four telescopes since 2004, either sparsely during the H.E.S.S. monitoring program, 
or intensely during dedicated
campaigns such as that described in \cite{HESS2155_MWL}~(2005c), showing mean flux levels of $\sim20\%$ of the Crab Nebula flux
for energies above $200\,{\rm GeV}$. During the summer of 2006, \pks\ exhibited unprecedented
flux levels accompanied by strong variability (\cite{HESS2155_BigFlare}~2007a), making temporal and
spectral variability studies possible on timescales of the order of a few minutes. The VHE
$\gamma$-ray emission is usually thought to originate from a relativistic jet, emanating from
the vicinity of a Supermassive Black Hole (SMBH). The physical processes at play are
still poorly understood, but the analysis of the $\gamma$-ray flux spectral and temporal
characteristics is well suited to provide better insights.  
 
For this goal the data set of H.E.S.S. observations of
\pks\ between 2005 and 2007 is used. After describing the observations and the analysis 
chain in Section~\ref{ObsAna},
the emission from the ``quiescent'', i.e. nonflaring, state of the source will be characterized in
Section~\ref{QuiescentState}.
Section~\ref{SpecVar} details spectral variability
related to the source intensity. Section~\ref{temporal_variab} will focus on the description of
the temporal variability during the highly active state of the source, and its possible energy
dependence. Section~\ref{log_normal_process} will illustrate a description of the observed
variability phenomenon by a random stationary process, characterized by a simple power density
spectrum. Section~\ref{characteristic_time} will show how limits on the characteristic time of 
the source can be derived.
The multi-wavelength aspects from the high flux state will be presented in a second paper.

\section{Observations and analysis}
\label{ObsAna}

H.E.S.S. is an array of four imaging atmospheric Cherenkov telescopes situated
in the Khomas Highland of Namibia ($23^\circ16\arcmin18\arcsec$ South,
$16^\circ30\arcmin00\arcsec$ East), at an elevation of 1,800 meters above sea
level (see \cite{HESS_Crab}~2006).  \pks\ was observed by H.E.S.S. each year since
2002; results of observations in 2002, 2003 and 2004 can be found in
\cite{HESS2155_2003}~(2005b), \cite{HESS2155_MWL}~(2005c) and \cite{HESS2155_Dust}.  The data
reported here were collected between 2005 and 2007.  In 2005, 12.2 hours of
observations were taken.  A similar observation time was scheduled in 2006, but
following the strong flare of July 26th (\cite{HESS2155_BigFlare}~2007a) it was
decided to increase this observation time significantly.
Ultimately, from June to October 2006,
this source was observed for 75.9 hours, with a further 20.9 hours in 2007.  

\begin{table}
\begin{center}
\caption{Summary of observations for each year. $T$ represents the live-time (hours), 
$n_\mathrm{on}$ the number of on-source events, $n_\mathrm{off}$ the number of off-source events 
(from a region five times larger than for the on-source events), 
and $\sigma$ the significance of the corresponding excess, given in units of standard deviations.
\label{JournalYears}}
\begin{tabular}{*{8}{c}}
   \hline
Year   & $T$   &   $n_\mathrm{on}$ &  $n_\mathrm{off}$ &   Excess&  $\sigma$ & $\sigma/\sqrt{T}$ \\ 
   \hline
2005 & 9.4 &  7,282 & 27,071 &  1,868 &  21.8 &   7.1 \\
2006 & 66.1 & 123,567 & 203,815 & 82,804 & 288.4 &  35.5 \\
2007 & 13.8 & 11,012 & 40,065 &  2,999 &  28.6 &   7.7 \\
   \hline
Total & 89.2 & 141,861 & 270,951 & 87,671 & 275.6 &  29.2 \\
   \hline
\end{tabular}
\end{center}
\end{table}

\begin{figure}[!t]
  \centering \includegraphics[width=9cm]{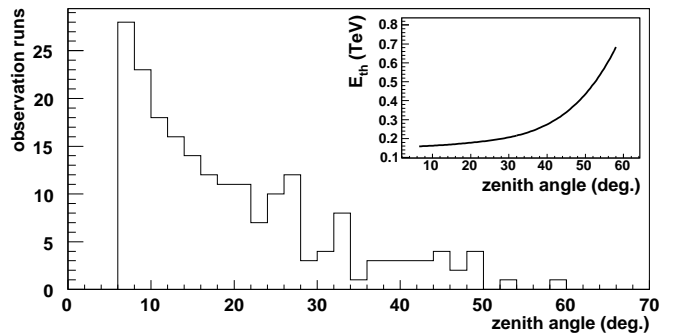}
  \caption{
Zenith angle distribution for the 202 4-telescopes observation runs from 2005 to 2007. 
The inset shows, for each zenith angle, the energy threshold associated with the analysis 
presented in Section~\ref{ObsAna}.
}
  \label{ZenDist}
\end{figure}

The data have been recorded during runs of 28 minutes nominal duration, 
with the telescopes pointing at $0.5^\circ$ from the source position in the sky
to enable a simultaneous estimate of the background. 
This offset has been taken alternatively in both right ascension and declination (with both signs), 
in order to minimize systematics. 
Only the runs passing the H.E.S.S. data-quality selection criteria 
have been used for the analyses presented below. 
These criteria imply good atmospheric conditions and 
checks that the hardware state of the cameras is satisfactory. 
The number of runs thus selected is 22 for 2005, 153 for 2006 and 35 for 2007, 
corresponding to live-times of 9.4, 66.1 and 13.8 hours respectively.
During these observations, zenith angles were between 7 and 60 degrees, 
resulting in large variations in the instrument energy threshold ($E_\mathrm{th}$, see Fig.~\ref{ZenDist}) and sensitivity. 
This variation has been accounted for in the spectral and temporal variability studies presented below.

The data have been analysed following the prescription presented in \cite{HESS_Crab}~(2006),
using the {\it loose} set of cuts which are well adapted for bright sources with moderately soft spectra,
and the {\it Reflected-Region} method for the definition of the on-source and off-source data regions.
A year-wise summary of the observations and the resulting detections are shown in Table \ref{JournalYears}.
A similar summary is given in Appendix \ref{JournalNights} for the 67 nights of data taken,
showing that the emission of \pks\ is easily detected by H.E.S.S. almost every night.
For 66 nights out of 67, the significance per square root of the live-time ($\sigma/\sqrt{T}$, where $T$ 
is the observation live-time) is at least equal to $3.6\,\sigma/\sqrt{h}$, 
the only night with a lower value --- MJD 53705 --- corresponding to a very short exposure.
In addition, for 61 nights out of 67 the source emission is high enough to enable a detection of the 
source with 5$\,\sigma$ significance in one hour or less, a level usually required in this domain to 
firmly claim a new source detection.
In 2006 the source exhibits very strong activity (38 nights, 
between MJD 53916 -- 53999) with a nightly $\sigma/\sqrt{T}$ varying from 3.6 to 150, 
and being higher than $10\,\sigma/\sqrt{h}$ for 19 nights. The activity of the 
source climaxes on MJD\,53944 and 53946 with statistical significances 
which are unprecedented at these energies, the rate of detected $\gamma$-rays corresponding to 
2.5 and $1.3\;\mathrm{Hz}$, with 150 and $98\,\sigma/\sqrt{h}$ respectively.

For subsequent spectral analysis, an improved energy reconstruction method with respect to the one 
described in \cite{HESS_Crab}~(2006) was applied. 
This method is based on a look-up table determined from Monte-Carlo simulations, which contains the relation 
between an image's amplitude and its reconstructed impact parameter as a function of the true energy, 
the observation zenith angle, the position of the source in the camera, the optical efficiency of the 
telescopes (which tend to decrease due to the aging of the optical surfaces), the number of triggered telescopes and the reconstructed 
altitude of the shower maximum. Thus, for a given event, the reconstructed energy is determined by 
requiring the minimal $\chi^2$ between the image amplitudes 
and those expected from the look-up table corresponding to the same observation conditions. 
This method yields a slightly lower energy threshold (shown in Fig.~\ref{ZenDist} as a function of zenith angle), 
an energy resolution which varies from 15\% to 20\% over all the energy range, and biases in the energy
reconstruction which are smaller than 5\%, 
even close to the threshold. The systematic uncertainty in the normalization
of the H.E.S.S. energy scale is estimated to be as large as 15\%, corresponding for such soft 
spectrum source to ~40\% in the overall flux normalization as quoted in \cite{HESS2155_Chandra}~(2009).

All the spectra presented in this paper have been obtained using a forward-folding maximum
likelihood method based on the measured energy-dependent on-source and off-source 
distributions. This method, fully described in \cite{CAT_Mrk421}, performs a global 
deconvolution of the instrument functions (energy resolution, collection area) and 
the parametrization of the spectral shape.
Two different sets of parameters, corresponding to a power law and 
to a power law with an exponential cut-off, are used for the spectral 
shape, with the following equations :
\begin{eqnarray}
\lefteqn{\phi(E) = \phi_0 \big(\frac{E}{E_0}\big)^{-\Gamma}}\label{E:eq1}\\
\lefteqn{\phi(E) = \phi_0 \exp\big({\frac{E_0}{E_\mathrm{cut}}}\big)\big(\frac{E}{E_0}\big)^{-\Gamma}\exp\big(-{\frac{E}{E_\mathrm{cut}}}\big)}\label{E:eq2}
\end{eqnarray}
$\phi_0$ represents the differential flux at $E_0$ (chosen to be 1~TeV), 
$\Gamma$ is the power law index and $E_\mathrm{cut}$ the characteristic energy of the exponential cut-off.
The maximum likelihood method provides the best set of parameters corresponding to the selected hypothesis, 
and the corresponding error matrix.

\begin{table*}[!]
\begin{center}
\caption{
The various data sets used in the paper, referred to in the text by the labels presented in this table.
Only runs with the full array of four telescopes in operation (202 runs over 210) and an 
energy threshold lower than 200\,GeV (165 runs over 202) are considered.
The corresponding period of the observations, the number of runs, the live-time $T$ (hours),
the number of $\gamma$ excess events and its significance $\sigma$ are shown. 
The column $section$ indicates the sections of the paper in which each data set is discussed. 
Additional criteria for the data set definitions are indicated in the last column.
\label{DataSetTab}}
\begin{tabular*}{0.79\textwidth}{*{8}{c}}
   \hline
Label&   Period   &Runs& $T$(hours)    &Excess &  $\sigma$  & Section  & Additional criteria \\ 
   \hline
$D$   & 2005--2007 &165 & 69.7 & 67,654 & 237.4 &  \ref{SpecVar}, \ref{characteristic_time}  & --  \\
$D_{QS}$   & 2005--2007 &115 & 48.1 & 12,287 &  60.5 &  \ref{QuiescentStateDistrib}, \ref{QuiescentStateSpectrum}, \ref{IntrinsicWidth}, \ref{characteristic_time} & July 2006 excluded \\
$D_{QS-2005}$  &    2005    & 19 &  8.0 &  1,816 &  22.6 &  \ref{QuiescentStateSpectrum} & -- \\
$D_{QS-2006}$  &    2006    & 61 & 26.3 &  7,472 &  48.4 &  \ref{QuiescentStateSpectrum} & July 2006 excluded \\
$D_{QS-2007}$  &    2007    & 35 & 13.8 &  2,999 &  28.6 &  \ref{QuiescentStateSpectrum} & -- \\
$D_{JULY06}$ &July 2006& 50 & 21.6 & 55,367 & 281.8 &  \ref{SpecVar}, \ref{temporal_variab}, \ref{log_normal_process}, \ref{characteristic_time} & -- \\
$D_{FLARES}$ &July 2006 (4 nights)&27& 11.8 & 46,036 & 284.1 &  \ref{SpecVar}, \ref{temporal_variab}, \ref{log_normal_process}, \ref{characteristic_time} & --\\
   \hline
\end{tabular*}
\end{center}
\end{table*}

Finally, various data sets have been used for subsequent analyses. These are summarized in Table~\ref{DataSetTab}.

\section{Characterization of the quiescent state}
\label{QuiescentState}

\begin{figure}
\centering 
\includegraphics[width=9cm]{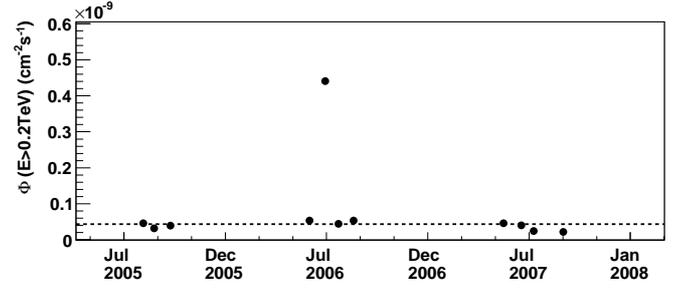}
\caption{
Monthly averaged integral flux of \pks\ above $200\,{\rm GeV}$ obtained from data set $D$
(see Table~\ref{DataSetTab}), which corresponds
to the 165 4-telescope runs whose energy threshold is below $200\,{\rm GeV}$. 
The dotted line corresponds to 15\% of the Crab Nebula emission level (see Section~\ref{QuiescentStateDistrib}).
\label{MonthLC}}
\end{figure}

As can be seen in Fig.~\ref{MonthLC}, with the exception of the high state of July 2006
\pks\ was in a low state during the observations from 2005 to 2007.
This section explores the variability of the source during these periods 
of low-level activity, based on the determination of the run-wise integral
fluxes for the data set $D_{QS}$, which excludes the flaring 
period of July 2006 and also those runs whose energy threshold is higher than $200\,{\rm GeV}$ (see~\ref{LCMethod} for justification).
As for Sections~\ref{temporal_variab} and ~\ref{log_normal_process}, 
the control of systematics in such a study is particularly important, especially
because of the strong variations of the energy threshold throughout the observations.

\subsection{Method and systematics}
\label{LCMethod}

The integral flux for a given period of observations is determined in a standard way.
For subsequent discussion purposes, the formula applied is given here :
\begin{eqnarray}
\label{intflxeq}
\Phi = N_\mathrm{exp}\frac{\int_{E_\mathrm{min}}^{E_\mathrm{max}}S(E)\mathrm{d}E}
       {T \int_{0}^{\infty}\int_{E_\mathrm{min}}^{E_\mathrm{max}} A(E)R(E,E')S(E)\mathrm{d}E'\mathrm{d}E}
\end{eqnarray}
where $T$ represents the corresponding live-time, $A(E)$ and $R(E,E')$ are, 
respectively, the collection area at the true energy $E$ and the 
energy resolution function between $E$ and the measured energy $E'$, and $S(E)$
the shape of the differential energy spectrum as defined in Eq.~\ref{E:eq1} and~\ref{E:eq2} .
Finally, $N_\mathrm{exp}$ is the number of measured events in the energy range 
[E$_\mathrm{min}$,E$_\mathrm{max}$].

In the case that $S(E)$ is a power law, an important source of systematic error in the determination
of the integral flux variation with time comes from the value chosen for the index $\Gamma$.
The average 2005--2007 energy spectrum yields a very well determined power law index\footnote{
The average 2005--2007 energy spectrum yields a power law with a photon index $\Gamma=3.37\pm0.02_\mathrm{stat}$.
One should note that some curvature is observed at higher energies, resulting in a better spectral determination 
when the alternative hypothesis shown in (Eq.~\ref{E:eq2}) is used, yielding a harder index 
($\Gamma=3.05\pm0.05_\mathrm{stat}$) with an exponential 
cut-off at energy $E_\mathrm{cut}=1.76\pm0.27_\mathrm{stat}~\mathrm{TeV}$. 
This curved model is prefered at a $8.4\,\sigma$
level as compared to the power law hypothesis. However, the choice of the model has little effect on
the determination of the integral flux values above $200\,{\rm GeV}$, the integral
being dominated by the low-energy part of the energy spectrum.}.
However, in Section~\ref{SpecVar} it will be shown that this index varies depending on the flux
level of the source. Moreover, in some cases the energy spectrum of the source shows some curvature in
the TeV region, giving slight variations in the fitted power law index depending on the energy range used.

For runs whose energy threshold is lower than $E_\mathrm{min}$, a simulation performed under 
the observation conditions corresponding to the data shows that an index variation of 
$\Delta\Gamma=0.1$ implies a flux error at the level of $\Delta\Phi\sim1\%$, this relation
being quite linear up to $\Delta\Gamma\sim0.5$. However, this relation no longer holds when
the energy threshold is above $E_\mathrm{min}$, as the determination of $\Phi$ becomes
much more dependent on the choice of $\Gamma$. For this reason, only runs whose energy
threshold is lower than $E_\mathrm{min}$ will be kept for the subsequent light curves. 
The value of $E_\mathrm{min}$ is chosen as $200\,{\rm GeV}$, which is a compromise 
between a low value which maximises the excess numbers used for the flux determinations and a high value which
maximises the number of runs whose energy threshold is lower than $E_\mathrm{min}$.

\subsection{Run-wise distribution of the integral flux}
\label{QuiescentStateDistrib}

\begin{figure*}
\centering
\includegraphics[width=0.90 \textwidth]{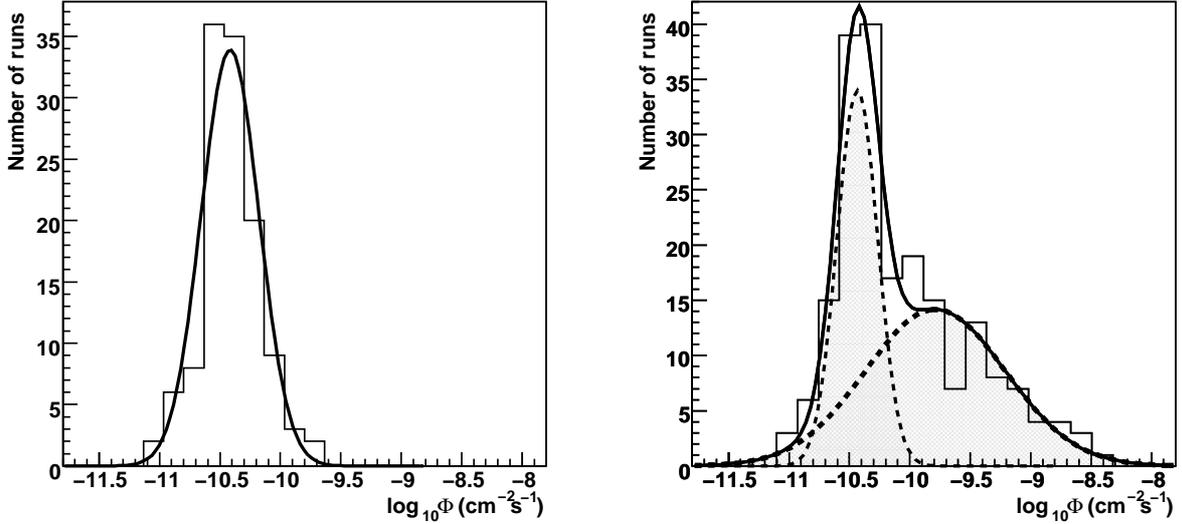}
\caption{Distributions of the logarithms of integral fluxes above $200\,{\rm GeV}$ in individual runs. 
Left: from 2005 to 2007 except the July 2006 period (data set $D_{QS}$), 
fitted by a Gaussian.
Right: all runs from 2005 to 2007 (data set $D$), where the solid line 
represents the result of a fit by the sum of 2 Gaussians (dashed lines).
See Table~\ref{tab:dbg} for details.}
\label{doubleGaussian_fit}
\end{figure*}

From 2005 to 2007, \pks\ is almost always detected when observed (except for two nights 
for which the exposure was very low), indicating
the existence, at least during these observations, of a minimal level of activity of the source.
Focussing on data set $D_{QS}$ (which excludes the July 2006 data where the source is in a high state),
the distribution of the integral fluxes of the individual runs above $200\,{\rm GeV}$ has been determined 
for the 115 runs, using a spectral index $\Gamma=3.53$ (the best value for this data set, 
as shown in~\ref{QuiescentStateSpectrum}).
This distribution has an asymmetric shape, with mean value 
$(4.32\pm0.09_\mathrm{stat}) \times 10^{-11}\,\percmsqrs$ and root mean square (r.m.s.) 
$(2.48\pm0.11_\mathrm{stat}) \times 10^{-11}\,\percmsqrs$, and is very well described with 
a lognormal function. Such a behaviour implies that the
logarithm of fluxes follows a normal distribution, centered on the
logarithm of $(3.75\pm0.11_\mathrm{stat}) \times 10^{-11}\,\percmsqrs$.
This is shown in the left panel of Fig.~\ref{doubleGaussian_fit}, where the solid line
represents the best fit obtained with a maximum-likelihood method, yielding results 
independent of the choice of the intervals in the histogram. It is interesting to note that this
result can be compared to the fluxes measured by H.E.S.S. from \pks\ during its construction 
phase, in 2002 and 2003 (see \cite{HESS2155_2003}~2005b and \cite{HESS2155_MWL}~2005c).  As shown in
Table~\ref{OldSpecVals}, these flux levels extrapolated down to $200\,{\rm GeV}$ 
were close to the value corresponding to the peak shown in the left panel of Fig.~\ref{doubleGaussian_fit}.

\begin{table}
\begin{center}
\caption{
Integral fluxes and their statistical errors from 2002 and 2003 
observations of \pks\ during the H.E.S.S. construction phase. 
These values are taken from \cite{HESS2155_2003}~(2005b) and \cite{HESS2155_MWL}~(2005c) and 
correspond to flux extrapolations to above $200\,{\rm GeV}$.\label{OldSpecVals}}
\begin{tabular}{*{3}{c}}
   \hline
Month  & Year  &  $\Phi$ [$10^{-11}\,\percmsqrs$]\\ 
   \hline
July & 2002 & $16.4\pm4.7$ \\
Oct. & 2002 &  $\,\,\,8.9\pm5.2$ \\
June & 2003 &  $\,\,\,5.8\pm1.4$ \\
July & 2003 &  $\,\,\,2.9\pm0.5$ \\
Aug. & 2003 &  $\,\,\,3.5\pm0.5$ \\
Sep. & 2003 &  $\,\,\,4.9\pm1.2$ \\
Oct. & 2003 &  $\,\,\,5.2\pm0.5$ \\
   \hline
\end{tabular}
\end{center}
\end{table}

The right panel of Fig.~\ref{doubleGaussian_fit} shows how the
flux distribution is modified when the July 2006 data are taken into account  
(data set $D$ in Table~\ref{DataSetTab}): the histogram 
can be accounted for by the superposition of two Gaussian distributions (solid curve).
The results, summarized in Table~\ref{tab:dbg}, are also independent
of the choice of the intervals in the histogram. 
Remarkably enough, the characteristics of the first Gaussian obtained in the first step
(left panel) remain quite stable in the double Gaussian fit.

\begin{table}
\caption{The distribution of the flux logarithm. First column: distribution as fitted by a single Gaussian law for the 
``quiescent'' regime (data set $D_{QS}$ ). Second column: distribution fitted by two
Gaussian laws, one for the ``quiescent'' regime, the other for
the flaring regime (data set $D$). Decimal logarithms are quoted
to make the comparison with the left panel of Fig.~\ref{doubleGaussian_fit} easier and the
flux is expressed in cm$^{-2}$~s$^{-1}$. In the first line the average of fluxes is reported, 
while in the second line their r.m.s..}
\label{tab:dbg} \centering
\begin{tabular}{c c c}
\hline
 & ``Quiescent'' regime & Flaring regime \\ \hline
 $\big< \log_{10}\Phi \big> $ & -10.42 $\pm$ 0.02 & -9.79 $\pm$ 0.11
 \\
r.m.s. $\: \mbox{of} \: \log_{10}\Phi$   & 0.24 $\pm$ 0.02   & 0.58 $\pm$ 0.04   \\
\hline
\end{tabular}
\end{table}

This leads to two conclusions. First, the flux distribution of \pks\ is well described considering
a low state and a high state, for each of which the distribution of the logarithms of the fluxes follows
a Gaussian distribution. The characteristics of the lognormal flux distribution for the high state
are given in Sections~\ref{temporal_variab}, \ref{log_normal_process} and \ref{characteristic_time}. 
Secondly, \pks\ has a level of minimal 
activity which seems to be stable on a several-year time-scale. This state will henceforth be 
referred to as the ``quiescent state'' of the source.\\

\subsection{Width of the run-wise flux distribution}
\label{IntrinsicWidth}

In order to determine if the measured width of the flux distribution (left panel of Fig.~\ref{doubleGaussian_fit}) can be explained
 as statistical fluctuations from the measurement process a simulation has been carried out considering a source 
which emits an integral flux above $200\,{\rm GeV}$ of 
$4.32 \times 10^{-11}\,\percmsqrs$ with a power law photon spectrum index 
$\Gamma=3.53$ (as determined in the next section). 
For each run of the data set $D_{QS}$ the number $n_{\gamma}$ 
expected by convolving the assumed differential energy spectrum with the instrument response
corresponding to the observation conditions is determined. 
A random smearing around this value allows statistical fluctuations to be taken into account. 
The number of events in the off-source region and also the number of background events in the 
source region are derived from the measured values $n_\mathrm{off}$
in the data set. These are also smeared in order to take into account the expected 
statistical fluctuations.

\begin{figure}
\centering
\includegraphics[width=9cm]{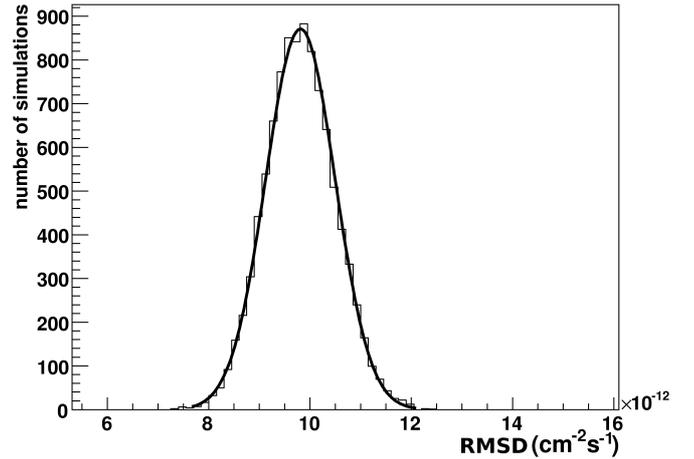}
\caption{
Distribution of RMSD obtained when the instrument response to a fixed emission ($\Phi=4.32\times 10^{-11}\,\percmsqrs$ and $\Gamma=3.53$)
is simulated 10000 times with the same observation conditions as for the 115 runs of the left part of Fig.~\ref{doubleGaussian_fit}.
\label{SimDistrib}
}
\end{figure}

10000 such flux distributions have been simulated, and for each one its mean value and 
r.m.s. (which will be called below RMSD) are determined. The distribution of RMSD thus obtained, shown in
Fig.~\ref{SimDistrib}, is well described by a Gaussian centred on $0.98\times 10^{-11}\,\percmsqrs$ 
(which represents a relative flux dispersion of 23\%) and with a $\sigma_{RMSD}$ 
of $0.07\times 10^{-11}\,\percmsqrs$. 

It should be noted that here the effect of atmospheric fluctuations in the determination of the flux is only taken 
into account at the level of the off-source events, as these numbers are taken from the measured data. 
But the effect of the corresponding level of fluctuations on the source signal is very difficult to determine. 
If a conservative value of 20\% is considered
\footnote{
\label{fn_crabatm}
A similar procedure has been carried out on the Crab Nebula observations. Considering this source to be perfectly stable 
it allows us to determine an upper limit to the fluctuations of the Crab signal
due to the atmosphere, and a value of $\sim10\%$ was derived. Nonetheless, this value is linked to the
observations' epoch and zenith angles, and to the source spectral shape.
}, which is added in the simulations
as a supplementary fluctuation factor for the number of events expected from the source,
a RMSD distribution centred on $1.30\times 10^{-11}\,\percmsqrs$ with a $\sigma_{RMSD}$ 
of $0.09\times 10^{-11}\,\percmsqrs$ is obtained.
Even in this conservative case, the measured value for the flux distribution r.m.s. 
($(2.48\pm0.11_\mathrm{stat}) \times 10^{-11}\,\percmsqrs$) is very far (more than 8 standard deviations) 
from the simulated value. 
All these elements strongly suggest the existence of an intrinsic variability associated with the quiescent state of \pks.

\subsection{Quiescent-state energy spectrum}
\label{QuiescentStateSpectrum}

\begin{figure}[h!]
\centering 
\includegraphics[width=9cm]{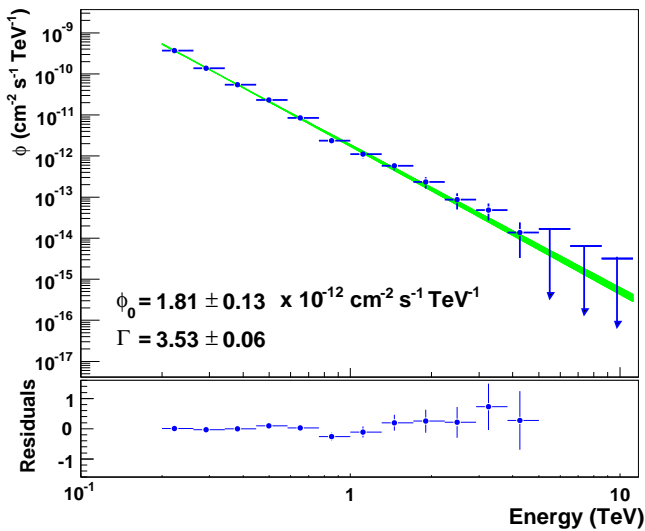}
\caption{
\label{SpecSteadyState}
Energy spectrum of the quiescent state for the period 2005-2007.
The green band correponds to the 68\% confidence-level provided by the maximum likelihood method.
Points are derived from the residuals in each energy bin, only for illustration purposes.
See Section~\ref{QuiescentStateSpectrum} for further details.}
\end{figure}

The energy spectrum associated with the data set $D_{QS}$, shown in Fig.~\ref{SpecSteadyState}, 
is well described by a power law with 
a differential flux at 1\,TeV of
$\phi_0=(1.81\pm0.13_\mathrm{stat}) \times 10^{-12}\,\percmsqrsT$ and an index of $\Gamma=3.53\pm0.06_\mathrm{stat}$. 
The stability of these values for spectra measured separately for 2005, 2006 (excluding July) and 2007 
is presented in Table~\ref{SpecQSYear}.
The corresponding average integral flux is $(4.23\pm0.09_\mathrm{stat}) \times 10^{-11}\,\percmsqrs$,
which is as expected in very good agreement with the mean value of the distribution shown in the left
panel of Fig.~\ref{doubleGaussian_fit}.

\begin{table}
\begin{center} 
\caption{
Parametrization of the differential energy spectrum of the quiescent state of \pks, 
determined in the energy range 0.2--10\,TeV, first for the
2005--2007 period and also separately for the 2005, 2006 (excluding July) and 2007 periods. 
Corresponding data sets are those of Table~\ref{DataSetTab}.
$\phi_0$ ($10^{-12}\,\percmsqrsT$) is the differential flux at 1\,TeV, $\Gamma$ the photon index and 
$\Phi$ ($10^{-11}\,\percmsqrs$) the integral flux above 0.2\,TeV. Errors are statistical.
\label{SpecQSYear}
}
\begin{tabular}{*{5}{c}}
   \hline
Year  & Data set & $\phi_0$ &  $\Gamma$ & $\Phi$ \\ 
   \hline

2005--2007 & $D_{QS}$  &$1.81\pm0.13$ & $3.53\pm0.06$ & $4.23\pm0.09$ \\
      2005 & $D_{QS-2005}$ &$1.59\pm0.32$ & $3.56\pm0.16$ & $3.83\pm0.21$\\
      2006 & $D_{QS-2006}$ &$1.87\pm0.18$ & $3.59\pm0.08$ & $4.65\pm0.13$\\
      2007 & $D_{QS-2007}$ &$1.84\pm0.24$ & $3.43\pm0.11$ & $3.78\pm0.16$\\
   \hline
\end{tabular}
\end{center}
\end{table}
Bins above 2\,TeV correspond to $\gamma-$ray excesses lower than 20$\,\gamma$ and significances lower than $2\,\sigma$.
Above 5\,TeV excesses are even less significant ($\sim1\sigma$ or less) and 99\% upper-limits are used.
There is no improvement of the fit when a curvature is taken into account.


\section{Spectral variability}\label{spectral_variab}
\label{SpecVar}


\subsection{Variation of the spectral index for the whole data set 2005-2007}
\label{SpecVarAll}


The spectral state of \pks\ has been monitored since 2002.  
The first set of observations (\cite{HESS2155_2003}~2005b), from July 2002 to September 2003,
shows an average energy spectrum well described by a power law with an index
of $\Gamma=3.32\pm0.06_\mathrm{stat}$, for an integral flux (extrapolated down to $200\,{\rm GeV}$
) of $(4.39\pm0.40_\mathrm{stat})\times 10^{-11}\,\percmsqrs$. No clear indication of
spectral variability was seen. Consecutive observations in October and
November 2003 (\cite{HESS2155_MWL}~2005c) gave a similar value for
the index, $\Gamma=3.37\pm0.07_\mathrm{stat}$, for a slightly higher flux of
$(5.22\pm0.54_\mathrm{stat})\times 10^{-11}\,\percmsqrs$.  
Later, during H.E.S.S. observations of the first (MJD\,53944, \cite{HESS2155_BigFlare}~2007a) 
and second (MJD\,53946, \cite{HESS2155_Chandra}~2009) exceptional flares of July 2006, 
the source reached much higher average fluxes, corresponding to 
$(1.72\pm0.05_\mathrm{stat})\times 10^{-9}\,\percmsqrs$ and $(1.24\pm0.02_\mathrm{stat})\times 10^{-9}\,\percmsqrs$
\footnote{corresponding to data set T200 in \cite{HESS2155_Chandra}~(2009)} respectively.
In the first case, no strong indications for spectral variability were found and the average index 
$\Gamma=3.19\pm0.02_\mathrm{stat}$ was close to those associated with the 2002 and 2003 observations. 
In the second case, clear evidence of spectral hardening with increasing flux was found.

\begin{figure}[!]
  \centering 
  \includegraphics[width=9cm]{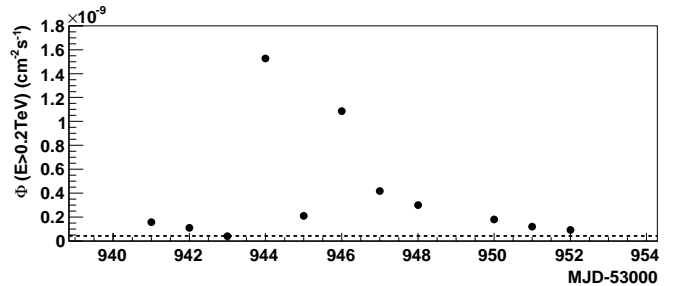}
  \caption{Integral flux above $200\,{\rm GeV}$ measured each night during late July 2006 observations. 
The horizontal dashed line corresponds to the quiescent state emission level defined in 
Section~\ref{QuiescentStateDistrib}.
\label{JulyNightsLC}}
\end{figure}

The observations of \pks\ presented in this paper also include 
the subsequent flares of 2006 and the data of 2005 and 2007. 
Hence, the evolution of the spectral index is studied for the first time for a flux level varying 
over two orders of magnitude.
This spectral study has been carried out over the fixed energy range 0.2--1\,TeV in order to minimize both 
systematic effects due to the energy threshold variation and the effect of the curvature observed 
at high energy in the flaring states.
The maximal energy has been chosen to be at the limit where the spectral curvature seen in high flux 
states begins to render the power law or exponential curvature hypotheses distinguishable.
As flux levels observed in July 2006 are significantly higher than in the rest of the data set 
(see Fig.~\ref{JulyNightsLC}), the flux--index behaviour is determined separately first 
for the July 2006 data set itself ($D_{JULY06}$) and secondly for the 2005-2007 data excluding this data set
($D_{QS}$).

On both data sets, the following method was applied. The integral flux was determined 
for each run assuming a power law shape with an index of $\Gamma=3.37$ (average spectral index for the whole data set), 
and runs were sorted by increasing flux. The set of ordered runs was then divided into subsets containing at least an excess 
of 1500 $\gamma$ above $200\,{\rm GeV}$ and the energy spectrum of each subset was determined\footnote{even for lower fluxes, the significance associated with each subset is always higher than 20 standard deviations}.

\begin{figure*}[!]
  \centering 
  \includegraphics[angle=0,scale=0.9]{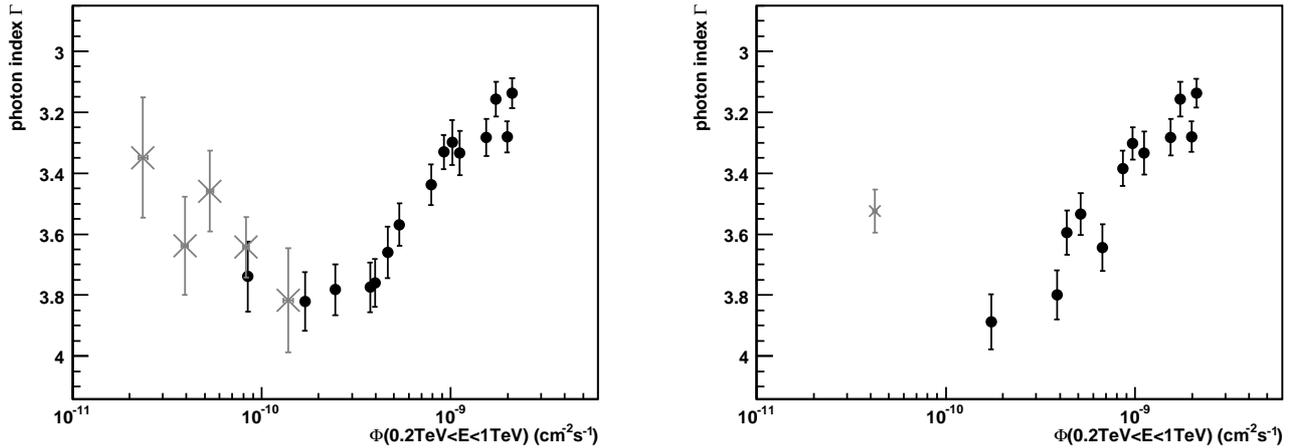}
  \caption{
Evolution of the photon index $\Gamma$ with increasing flux $\Phi$ in the 0.2--1\,TeV energy range.
The left panel shows the results for the July 2006 data (black points, data set $D_{JULY06}$) and for 
the 2005--2007 period excluding July 2006 (grey points, data set $D_{QS}$). 
The right panel shows the results for the four nights flaring period of July 2006 (black points, data set $D_{FLARES}$)
and one point corresponding to the quiescent state average spectrum (grey point, again data set $D_{QS}$).
See text in Sections  \ref{SpecVarAll} (left panel) and \ref{SpecFlaringNights} (right panel) 
for further details on the method.
\label{SpecVar20052007_mix}
}
\end{figure*}

The left panel of Fig.~\ref{SpecVar20052007_mix} shows the photon index versus integral flux for
data sets $D_{QS}$ (grey crosses) and $D_{JULY06}$ (black points).
Corresponding numbers are summarized in Appendix~\ref{SpecVarApp}.
While a clear hardening is observed for integral fluxes above a few $10^{-10}\,\percmsqrs$, 
a break in this behaviour is observed for lower fluxes. 
Indeed, for the data set $D_{JULY06}$ (black points) a linear fit yields a slope 
$\rm d\Gamma$/$\rm d\Phi=(3.0\pm0.3_\mathrm{stat}) \times 10^{8}\ensuremath{\mathrm{cm}^{2}\mathrm{s}}$,
whereas the same fit for data set $D_{QS}$ (grey crosses) yields  
a slope $\rm d\Gamma$/$\rm d\Phi=(-3.4\pm1.9_\mathrm{stat}) \times 10^{9}\ensuremath{\mathrm{cm}^{2}\mathrm{s}}$.
The latter corresponds to a $\chi^2$ probability $\rm P(\chi^2)=71\%$; 
a fit to a constant yields $\rm P(\chi^2)=33\%$ but with a constant fitted index incompatible 
with a linear extrapolation from higher flux states at a 3$\,\sigma$ level.
This is compatible with conclusions obtained either with an independant analysis or when these spectra are 
processed following a different prescription. In this prescription the runs were sorted
as a function of time in contiguous subsets with similar photon statistics, rather than
as a function of increasing flux.

The form of the relation between the index versus integral flux is unprecedented in the TeV regime. 
Prior to the results presented here, spectral variability has been detected only in two
other blazars, Mrk\,421 and Mrk\,501. For Mrk\,421, a clear hardening with increasing flux appeared 
during the 1999/2000 and 2000/2001 observations performed with HEGRA (\cite{Mrk421_HEGRA}) 
and also during the 2004 observations performed with H.E.S.S. (\cite{Mrk421_HESS}). 
In addition, the Mrk\,501 observations carried out with CAT during the 
strong flares of 1997 (\cite{CAT_Mrk501}) and also the recent observation 
performed by MAGIC in 2005 (\cite{magic_mrk501}~2007) have shown a similar hardening. 
In both studies, the VHE peak has been observed in the $\nu F_\nu$ distributions of 
the flaring states of Mrk\,501.

\subsection{Variation of the spectral index for the four flaring nights of July 2006}
\label{SpecFlaringNights}
\begin{figure*}[!]
  \centering 
  \includegraphics[angle=0,scale=1.02]{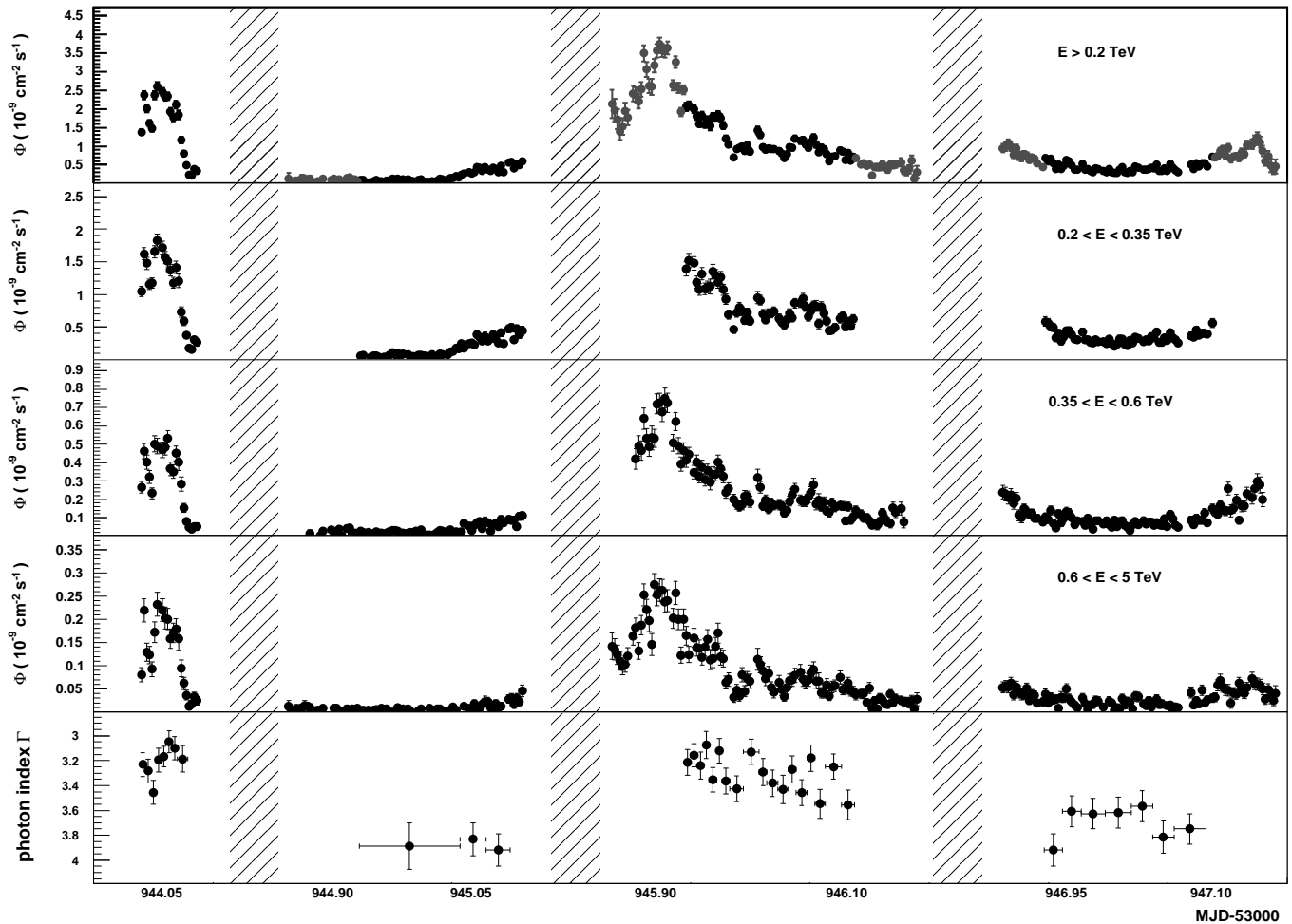}
  \caption{ Integrated flux versus time for \pks\ on MJD\,53944--53947 for four
    energy bands and with a 4-minute binning. From top to bottom: $>0.2\,{\rm
      TeV}$, 0.2--$0.35\,{\rm TeV}$, 0.35--$0.6\,{\rm TeV}$ and 0.6--$5\,{\rm
      TeV}$.  These light curves are obtained using a power law spectral shape
    with an index of $\Gamma=3.37$, also used to derive the flux extrapolation down to
    $0.2\,{\rm TeV}$ when the threshold is above that energy in the top panel
    (grey points).  Because of the high dispersion of the energy threshold of
    the instrument (see Section~\ref{ObsAna}, Fig.~\ref{ZenDist}), and following the prescription described
    in~\ref{LCMethod}, the integral flux has been determined for a time bin only
    if the corresponding energy threshold is lower than $E_\mathrm{min}$.  The
    fractional r.m.s. for the light curves are respectively, $0.86\pm0.01_\mathrm{stat}$,
    $0.79\pm0.01_\mathrm{stat}$, $0.89\pm0.01_\mathrm{stat}$ and $1.01\pm0.02_\mathrm{stat}$. 
    The last plot shows the variation of the photon index determined in the 0.2--1 TeV range. 
See Section~\ref{SpecFlaringNights} and Appendix~\ref{SpecVar4nightsTab}
 for details.}
\label{FourNightsLC}
\end{figure*}

In this section, the spectral variability during the flares of July 2006 is described in more detail.
A zoom on the variation of the integral flux (4-minute binning) for the four nights containing 
the flares (nights MJD\,53944, 53945, 53946, and 53947, called the ``flaring period'')
is presented in the top panel of Fig.~\ref{FourNightsLC}. This figure shows two exceptional 
peaks on MJD\,53944 and MJD\,53946 which climax respectively at fluxes higher than $2.5\times 10^{-9}\,\percmsqrs$
and $3.5\times 10^{-9}\,\percmsqrs$ ($\sim9$ and $\sim12$ times the Crab Nebula level above the
same energy), both about two orders of magnitude above the quiescent state level.

The variation with time of the photon index is shown in the bottom panel of Fig.~\ref{FourNightsLC}.
To obtain these values, the $\gamma$ excess above $200\,{\rm GeV}$ has been determined for each 4-minute bin.
Then, successive bins have been grouped in order to reach a global excess higher than 600 $\gamma$. 
Finally, the energy spectrum of each data set has been determined in the 
0.2--1 TeV energy range, as before (corresponding numbers are summarized in Appendix 
Table~\ref{SpecVar4nightsTab}).
There is no clear indication of spectral variability within each night, except for MJD\,53946 as 
shown in \cite{HESS2155_Chandra}~(2009). However, a variability can be seen from night to night, 
and the spectral hardening with increasing flux level already shown in Fig.~\ref{SpecVar20052007_mix} 
is also seen very clearly in this manner.

It is certainly interesting to directly compare the spectral behaviour seen during the flaring period with
the hardness of the energy spectrum associated with the quiescent state. This is shown in the right
panel of Fig.~\ref{SpecVar20052007_mix}, where black points correspond to the four flaring nights; these were
determined in the same manner as for the left panel (see \ref{SpecVarAll} for details). A linear fit here yields
a slope $\rm d\Gamma$/$\rm d\Phi=(2.8\pm0.3_\mathrm{stat}) \times 10^{8}\ensuremath{\mathrm{cm}^{2}\mathrm{s}}$.
The grey cross corresponds to the integral flux and the photon index associated with the quiescent state (derived 
in a consistent way in the energy range from 0.2--1 TeV), showing a clear rupture with the tendancy at 
higher fluxes (typically above $10^{-10}\,\percmsqrs$).

These four nights were further examined to search for differences in the spectral
behaviour between periods in which the source flux was clearly increasing and periods
in which it was decreasing.
For this, the first 16 minutes of the first flare (MJD\,53944) are of special interest
because they present a very symmetric situation: the flux increases during the first half, 
and then decreases to its initial level, the averaged fluxes are similar in both parts 
($\sim1.8\times 10^{-9}\,\percmsqrs$), and the observation conditions (and thus the instrument response) 
are almost constant --- the mean zenith angle of each part being respectively $7.2$ and $7.8$ degrees.
Again, the spectra have been determined in the 0.2--1\,TeV energy range, 
giving indices of $\Gamma=3.27\pm0.12_\mathrm{stat}$ and $\Gamma=3.27\pm0.09_\mathrm{stat}$ respectively.
To further investigate this question and avoid potential systematic errors from the spectral method determination,
the hardness ratios were derived (defined as the ratio of the excesses in different energy bands), using
for this the energy (TeV) bands [0.2--0.35], [0.35--0.6] and [0.6--5.0]. For any combination, 
no differences were found beyond the $1\,\sigma$ level between the increasing and decreasing 
parts.
A similar approach has been applied --- when possible --- for the rest of the flaring period.
No clear dependence has been found within the statistical error limit of the determined indices, 
which is distributed between 0.09 and 0.20.

Finally, the persistence of the energy cut-off in the differential energy spectrum along the flaring period
has been examined. For this purpose, runs were sorted again by increasing flux and grouped into subsets 
containing at least an excess of 3000 $\gamma$
above $200\,{\rm GeV}$\footnote{To be significant,
the determination of an energy cut-off needs a greater number of $\gamma$ than for a power law fit.}. 
For the seven subsets found, the energy spectrum has been determined in the 0.2--10\,TeV energy range 
both for a simple power 
law and a power law with an exponential cut-off. This last hypothesis was found to be favoured systematically 
at a level varying from 1.8 to 4.6 $\sigma$ compared to the simple power-law and is always compatible with
a cut-off in the 1--2 TeV range.

\section{Light curve variability and correlation studies}\label{temporal_variab}

This section is devoted to the characterization of the temporal variability of \pks, focusing
on the flaring period observations. The high number of $\gamma$-rays available not only allowed
minute-level time scale studies, such as those presented for MJD\,53944 in
\cite{HESS2155_BigFlare}~(2007a), but also to derive detailed light curves for three energy bands (Fig.~\ref{FourNightsLC}): 0.2--0.35\,TeV,
0.35--0.6\,TeV and 0.6--5\,TeV. \\ 
The variability of the energy-dependent light curves of \pks\ is in the following quantified through
their fractional r.m.s. $F_{\rm var}$ defined in Eq.~\ref{eq_fvar} (\cite{nandra,edelson2002}). In addition, possible time lags between light
curves in two energy bands are investigated.


\subsection{Fractional r.m.s. $F_{\rm var}$}\label{fvar}

All fluxes in the energy bands of Fig.~\ref{FourNightsLC} show a 
strong variability which is quantified through their fractional r.m.s. $F_{\rm var}$ (which depends on observation durations and their sampling). Measurement errors
$\sigma_{i,{\rm err}}$ on each of the $N$ fluxes $\phi_i$ of the light curve are taken into account in the definition of $F_{\rm var}$:

\begin{equation}\label{eq_fvar}
F_{\rm var}=\frac {\sqrt {S^2 - \sigma^{2}_{\rm err} }}{\overline{\phi}}
\end{equation}
where $S^2$ is the variance
\begin{equation}\label{eq_s2}
S^2=\frac{1}{N-1}\sum_{i=1}^{N}(\phi_i-\overline{\phi})^2,
\label{variance}
\end{equation}
and where $\sigma^2_\mathrm{err}$ is the mean square error and $\overline{\phi}$ is the mean flux.

The energy-dependent variability $F_{\rm var}(E)$ has been calculated for
the flaring period according to Eq.~\ref{eq_fvar} in all three energy
bands. The uncertainties on $F_{\rm var}(E)$ have been estimated according to
the parametrization derived by \cite{vaughan03}~(2003), using a Monte Carlo approach which accounts for the measurement errors on the simulated light curves.

\begin{figure}[tbh]
\centering
\includegraphics[width=0.45 \textwidth]{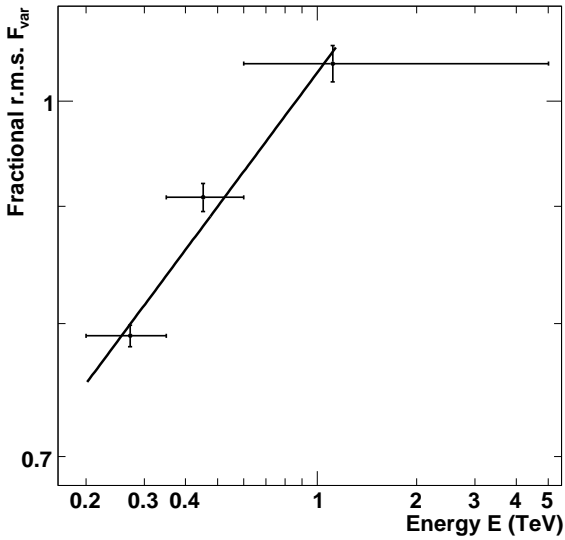}
\caption{Fractional r.m.s. $F_{\rm var}$ versus energy for the
  observation period MJD\,53944--53947. The points are the mean value
  of the energy in the range represented by the horizontal  bars. The line is
  the result of a power law fit where the errors on $F_{\rm var}$ and
  on the mean energy are taken into account, yielding $F_{\rm var}(E)\propto
  E^{0.19\pm0.01}$.}
\label{fvar_4nights}
\end{figure}

Fig.~\ref{fvar_4nights} shows the energy dependence of $F_{\rm var}$ over
the four nights for a sampling of 4~minutes where only fluxes with a
significance of at least 2 standard deviations were considered. There
is a clear energy-dependence of the variability (a null
probability of $\sim10^{-16}$). The points in Fig.~\ref{fvar_4nights}
are fitted according to a power law showing that the variability follows
$F_{\rm var}(E)\propto E^{0.19\pm0.01}$.\\

This energy dependence of $F_{\rm var}$ is also perceptible within each individual night. In Table~\ref{tab_fvar4nights} the values of $F_{\rm var}$, the relative mean flux, and
the observation duration, are reported night by night for the flaring
period. Because of the steeply falling
spectra, the low energy events dominate in the light
curves. This lack of statistics for high energy prevents to have a high fraction of points with a significance more than 2 standard deviation in light curves night by night for the three energy bands previously considered. On the other hand, the error contribution dominates and does not allow to estimate the $F_{\rm var}$ in all these three energy bands. Therefore, only two energy bands were considered: low (0.2--0.5\,TeV)
and high (0.5--5\,TeV). As can be seen in Table~\ref{tab_fvar4nights} also night by night the high-energy 
fluxes seem to be more variable than those at lower energies.

\begin{table}
\caption{Mean Flux and the fractional r.m.s. F$_{\rm var}$ night by night for MJD\,53944--53947. The values refers to light curves with 4 minute bins and respectively in three energy bands: $>$0.2\,TeV, 0.2--0.5\,TeV, 0.5--5.0\,TeV. Since a significant fraction ($\approx 40\%$) of the points in the light curve of MJD~53945 in the energy band 0.5--5.0\,TeV have a significance of less than 2 standard deviations, its $F_{\rm var}$ is not estimated.}
\label{tab_fvar4nights}      
\centering                          
\begin{tabular}{c c c c c}        
\hline                

MJD &Duration  & Energy& $\overline{\Phi}$ & $F_{\rm var}$ \\
 & (min) & (TeV) & (10$^{-10}$\,\percmsqrs) &  \\
\hline                        
53944 & 88  & & & \\
 & & all  & 15.44$\pm$0.87 &  0.56$\pm$0.01            \\
 & & 0.2 - 0.5&  13.28$\pm$0.85  &  0.55$\pm$0.01            \\
 & & 0.5 - 5.0&  1.94$\pm$0.24  & 0.61$\pm$0.03       \\
\hline
53945 & 244 & & & \\
 & & all  & 2.40$\pm$0.41 &  0.67$\pm$0.03            \\
 & & 0.2 - 0.5&  2.35$\pm$0.42  &  0.64$\pm$0.03            \\
 & & 0.5 - 5.0&  0.34$\pm$0.12  &     -     \\
\hline
53946 & 252 & & & \\
 & & all  & 11.39$\pm$0.80 &  0.35$\pm$0.01           \\
 & & 0.2 - 0.5&  10.02$\pm$0.79  &  0.33$\pm$0.01            \\
 & & 0.5 - 5.0&  1.39$\pm$0.20  &  0.43$\pm$0.02      \\
\hline
53947 & 252 & & & \\
 & & all  & 4.26$\pm$0.52 &  0.22$\pm$0.02           \\
 & & 0.2 - 0.5&  4.02$\pm$0.52  &  0.22$\pm$0.02            \\
 & & 0.5 - 5.0&  0.37$\pm$0.11  &  0.13$\pm$0.09      \\

\hline
                                 
\end{tabular}
\end{table}


\subsection{Doubling/halving timescale} \label{t2sect}

While $F_{\rm var}$ characterizes the mean variability of a source,
the shortest doubling/halving time (\cite{zhang1999}~1999) is an important
parameter in view of finding an upper limit on a possible physical shortest time scale of the
blazar. 

If $\Phi_i$ represents the light curve flux at a time $T_i$, for each
pair of $\Phi_i$ one may calculate $T_2^{i,j}=|\Phi \Delta T/ \Delta
\Phi|$, where $\Delta T$=$T_j$-$T_i$, $\Delta \Phi$=$\Phi_j$-$\Phi_i$
and $\Phi=(\Phi_j+\Phi_i)/2$. Two possible definitions of the
doubling/halving are proposed by \cite{zhang1999}~(1999): the smallest
doubling time of all data pairs in a light curve ($T_2$), or the mean
of the 5 smallest $T_2^{i,j}$ (in the following indicated as
$\tilde{T_2}$). One should keep in mind that, according 
to \cite{zhang1999}~(1999), these quantities are ill defined and strongly depend on the length of the sampling intervals and on the signal-to-noise ratio in the observation. \\ This quantity was calculated for the two nights with the highest fluxes, 
MJD\,53944 and MJD\,53946, considering light curves with two
different binnings (1  and 2 minutes). Bins with flux significances more than $2\,\sigma$ and flux ratios with an uncertainty smaller than 30\% were 
required to estimate the doubling time scale. The uncertainty on $T_2$
was estimated by propagating the errors on the $\Phi_i$, and a dispersion of the 5 smallest values was included in the error for $\tilde{T_2}$.

In Table~\ref{tab_t2}, the values of ${T_2}$ and $\tilde{T_2}$ for the two nights are
shown. The dependence with respect to the binning is clearly visible for both
observables. In this table, the last column shows that the fraction of pairs in the light curves which are kept in order to estimate the doubling times is on average $\sim$45\%. Moreover, doubling times $T_2$ and $\tilde{T_2}$ have been estimated for two sets of pairs in the light curves where $\Delta \Phi$=$\Phi_j$-$\Phi_i$ is increasing or decreasing respectively. The values of the doubling time for the two cases are compatible within $1,\sigma$, therefore no significant asymmetry has been found.      


\begin{table}
\caption{Doubling/Halving times for the high intensity nights  MJD\,53944 and 
  MJD\,53946 estimated with two different samplings, using
  the two definitions explained in the text. The final column corresponds to the fraction of flux pairs kept to estimate the doubling times.}          
\label{tab_t2}      
\centering 
\begin{tabular}{c c c c c }   
\hline           
 MJD & Bin size & $T_2$[min] & $\tilde{T_2}$[min] & Fraction of pairs  \\
\hline                      

53944 &  1~min & 1.65$\pm$0.38  & 2.27$\pm$0.77  &   0.53       \\
53944 &  2~min & 2.20$\pm$0.60  & 4.45$\pm$1.64   &  0.62       \\
\hline
53946  &  1~min & 1.61$\pm$0.45  & 5.72$\pm$3.83   &  0.25   \\
53946 & 2~min & 4.55$\pm$1.19  & 9.15$\pm$4.05     &  0.38   \\

\hline                 
\end{tabular}
\end{table}

 It should be noted that these values are strongly dependent on the time
 binning and on the experiment's sensitivity, so that the typical
 fastest doubling timescale should be conservatively estimated as being less than $\sim
 2\,{\rm min}$, which is compatible with the values reported in
\cite{HESS2155_BigFlare}~(2007a) and in \cite{magic_mrk501}~(2007), the latter concerning the blazar Mrk~501.

 
\subsection{Cross-correlation analysis as a function of energy}

Time lags between light curves at different energies can provide insights into acceleration, cooling and
propagation effects of the radiative particles. 

\begin{figure}[tbh]
\centering
\includegraphics[width=0.45 \textwidth]{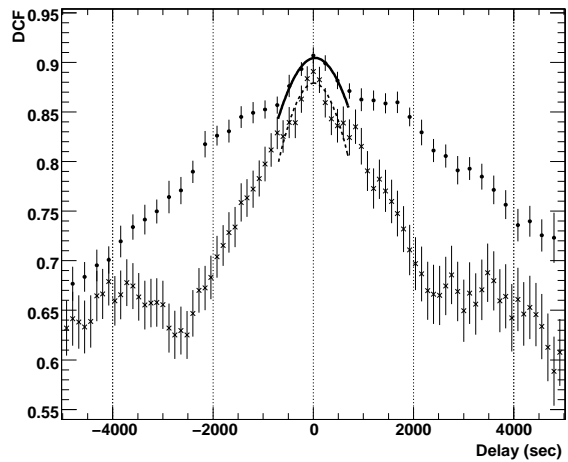} 
\caption{Discrete Correlation Function (DCF) between the light curves in the energy ranges 0.2-0.5 TeV
  and 0.5-5 TeV and Gaussian fits around the peak. Full circles
  represent the DCF for MJD\,53944--53947 4-minute light curve and the
  solid line is the Gaussian fit around the peak with mean value of $43\pm51\,{\rm s}$. Crosses represent the DCF for MJD\,53946
  with a 2-minute light curve binning, and the dashed line in the Gaussian fit with a
  peak centred at $14\pm41\,{\rm s}$.}
 \label{only_dcf}
\end{figure}

The Discrete Correlation Function (DCF) as a function of the delay
(\cite{white&peterson,edelson&krolik}) is used here to search for possible time lags
between the energy-resolved light curves. The uncertainty on the DCF
has been estimated using simulations. For each delay, $10^5$ light
curves (in both energy bands) have been generated within their errors,
assuming a Gaussian probability distribution. A probability
distribution function (PDF) of the correlation coefficients between
the two energy bands has been estimated for each set of simulated
light curves. The r.m.s. of these PDF are the errors related to the
DCF at each delay. Fig.~\ref{only_dcf} shows the DCF between the high
and low energy bands for the four-night flaring period (with 4 minute
bins) and for the second flaring night (with 2 minute bins). The gaps
between each 28 minute run have been taken into account in the DCF
estimation.

The position of the maximum of the DCF has been estimated by a
Gaussian fit, which shows no time lag between low and high energies
for either the 4 or 2 minute binned light curves. This sets a limit of $14\pm41\,{\rm s}$ from the observation of MJD\,53946. A detailed
study on the limit on the energy scale on which quantum gravity
effects could become important, using the same data set, are reported in
\cite{QG_hesspaper}~(2008a).

\subsection{Excess r.m.s.--flux correlation} \label{rmssect}

Having defined the shortest variability time scales, the nature of the process
which generates the fluctuations is investigated, using another estimator: the excess r.m.s.. It is defined as the variance of a light curve (Eq.~\ref{eq_s2}) after subtracting the measurement error:
\begin{equation} \label{excrms}
\sigma_\mathrm{xs}= {\sqrt{(S^2-\sigma^2_\mathrm{err})}}.
\end{equation}

\begin{figure}[!]
\centering
\includegraphics[width=0.5 \textwidth]{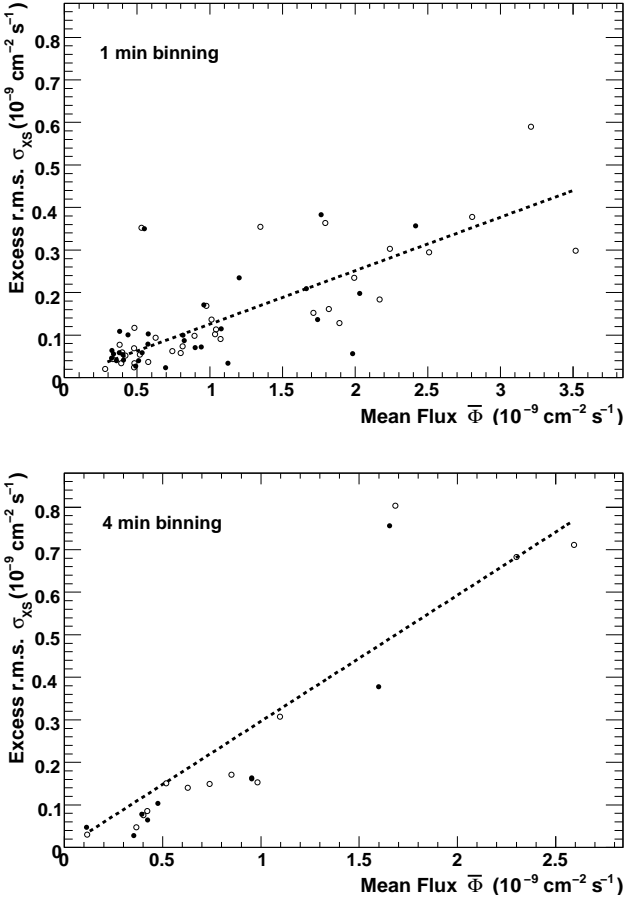} 
\caption{The excess r.m.s. $\sigma_{xs}$ {\em vs} mean flux $\bar{\Phi}$ for the observation in 
  MJD\,53944--53947 (Full circles). The open circles are the additional points obtained when also including the extrapolated flux points -- see text). Top: $\sigma_{xs}$ estimated with 20 minute time
  intervals and a 1 minute binned light curve. Bottom: $\sigma_{xs}$
  estimated with 80 minute time intervals and a 4 minute binned light
  curve. The dotted lines are a linear fit to the points, where
  $\sigma_{xs}\propto 0.13\times \bar{\Phi}$ for the 1 minute binning
  and $\sigma_{xs}\propto 0.3\times \bar{\Phi}$ for the 4 minute
  binning. Fits to the open circles yield similar results.} 
\label{rms-flux-exp}.
\end{figure}

Fig.~\ref{rms-flux-exp} shows the correlation between the excess r.m.s.
of the light curve and the flux, where the flux here considered are selected with an energy threshold of $200\,{\rm GeV}$.  
The excess variance is
estimated for 1- and 4-minute binned light curves, using 20 consecutive
flux points $\Phi_{i}$ which are at least at the $2\,\sigma$
significance level (81\% of the 1 minute binned sample). 
The correlation factors are
$r_1=0.60^{+0.21}_{-0.25}$ and $r_4=0.87^{+0.10}_{-0.24}$
for the 1 and 4 minute binning, excluding
an absence of correlation at the $2\,\sigma$ and $4\,\sigma$ levels
respectively, implying that fluctuations in the flux are probably proportional
to the flux itself which is a characteristic of lognormal
distributions (\cite{ait63}). This correlation has also been investigated extending the analysis to a 
statistically more significant data set including observations with a higher energy threshold in which the determination of the flux above $200\,{\rm GeV}$ requires an extrapolation (grey points in the top panel in Fig.~\ref{FourNightsLC}). In this case the correlations found are compatible ($r_1=0.78^{+0.12}_{-0.14}$ and $r_4=0.93^{+0.05}_{-0.15}$ for the 1 and 4 minute binning, respectively) and also exclude
an absence of correlation with a higher significance ($4\,\sigma$ and $7\,\sigma$, respectively).

Such a correlation has already been observed for X-rays in the Seyfert class AGN (\cite{edelson2002}, \cite{vf03}, \cite{vaughan03}~2003, \cite{mchardy04}) and in X-ray binaries (\cite{uttley01}, \cite{uttley04}, \cite{gleissner04}), where it is
considered as evidence for an underlying stochastic multiplicative
process (\cite{uttley05}), as opposed to an additive process. In additive processes, light
curves are considered as the sum of individual flares ``shots''
contributing from several zones (multi-zone models) and the relevant 
variable which has a Gaussian distribution (namely Gaussian
variable) is the flux. 
For multiplicative (or cascade) models
the Gaussian variable is the logarithm of the flux. Hence, this first observation of 
a strong r.m.s.-flux correlation in the VHE domain fully confirms the log-normality of the flux 
distribution presented in Section~\ref{QuiescentStateDistrib}.
\\

\section{Characterization of the lognormal process during the flaring period}\label{log_normal_process}

This section investigates whether the variability of \pks\ in the flaring period 
can be described by a random stationary process, where, as shown in Section~\ref{rmssect}, 
the Gaussian variable is the logarithm of the flux. 
In this case the variability can be characterized through its Power Spectral
Density (PSD) (\cite{vanderklis}) which indicates the density of
variance as function of the frequency $\nu$.
The PSD is an intrinsic indicator of the variability, usually
represented in large frequency intervals by power laws
($\propto\nu^{-\alpha}$) and is often used to define different
``states'' of variable objects (see e.g., \cite{paltani} and
\cite{zhang1999}~1999 for the PSD of \pks\ in the optical and
X-rays). The PSD of the light curve of one single night (MJD\,53944) was 
given in \cite{HESS2155_BigFlare}~(2007a) between $10^{-4}$ and $10^{-2}\,\mathrm{Hz}$, 
and was found to be compatible with a red noise process
($\alpha\geq$2) with $\sim10$ times more power as
in archival X-ray data (\cite{zhang1999}~1999), but with a similar index. 
This study implicitely assumed the $\gamma$-ray flux to be 
the Gaussian variable. In the present paper, the PSD is determined 
using data from 4 consecutive nights (MJD\,53944--53947) and 
assuming a lognormal process. Since direct Fourier analysis 
is not well adapted to light curves extending over multiple 
days and affected by uneven sampling and uneven flux errors, 
the PSD will be further determined on the basis of parametric 
estimation and simulations.


In the hypothesis where the process is stationary, i.e., the PSD is
time-independent, a power law shape of the PSD was assumed, as for
X-ray emitting blazars. The PSD  was defined as depending on two paramenters and as follows:
 $P(\nu) = K(\nu_{\rm ref}/
\nu)^{\alpha}$, where $\alpha$ is the variability spectral index and
$K$ denotes the ``power'' (i.e., the variance density) at a reference
frequency $\nu_{\rm ref}$. This latter was conventionally chosen to be $10^{-4}\,\mathrm{Hz}$, where the two parameters
$\alpha$ and $K$ are found to be decorrelated. Since a lognormal process is considered, $P(\nu)$ is the density of variance of the
Gaussian variable $\ln\,\Phi$. The 
\textit{natural} 
logarithm of the flux is
conveniently used here, since its variance over a given frequency
interval\footnote{If $\sigma^2$ is the variance of $\ln\,\Phi$,
$F_{\rm var} = \sqrt{\exp (\sigma^2)-1}$.} is close to the corresponding
value of $F_{\rm var}^2$, at least for small fluctuations. For a given
set of $\alpha$ and $K$, it is possible to simulate many long time
series, and to modify them according to experimental effects, 
namely those of background events and of flux measurement errors. 
Light curve segments are further extracted from this simulation, 
with exactly the same time structure (observation and non-observation
intervals) and the same sampling rates as those of real data. The
distributions of several observables obtained from simulations for
different values of $\alpha$ and $K$ will be compared to those determined
from observations, thus allowing these parameters to be determined 
from a maximum-likelihood fit.\\
The simulation characteristics will be described in
Section~\ref{simsec}. Sections~\ref{exc_sim}, \ref{str_fun} and
\ref{dbl} will be devoted to the determination of $\alpha$ and $K$
by three methods, each of them based on an experimental result: the
excess r.m.s.--flux correlation, the Kolmogorov first-order structure
function (\cite{rutman,simonetti}) and doubling-time measurements.

\subsection{Simulation of realistic time-series}\label{simsec}

For practical reasons, simulated values of $\ln\Phi$ were
calculated from Fourier series, thus with a discrete PSD. The
fundamental frequency $\nu_{0}=1/T_{0}$, which corresponds to an
elementary bin $\delta \nu \equiv \nu_0$ in frequency, must be much
lower than $1/T$ if $T$ is the duration of the observation. The ratio
$T_{0}/T$ was chosen to be of the order of 100, in such a way that
the influence of a finite value of $T_0$ on the average variance of
a light curve of duration $T$ would be less than about 2\%. Taking $T_0 =
9 \times 10^5\,\mathrm{s}$, this condition is fulfilled for the following
studies. With this approximation, the simulated flux logarithms are
given by:
\begin{equation}
\ln\,\Phi(t)=a_{0}+\sum_{n=1}^{N_\mathrm{max}}a_{n}\,\cos(2n\pi\nu_{0}t+\varphi_{n})
\end{equation}
where $N_\mathrm{max}$ is chosen in such a way that $T_{0}/N_\mathrm{max}$ is less
than the time interval between consecutive measurements (i.e., the
sampling interval). According to the definition of a Gaussian random
process, the phases $\varphi_{n}$ are uniformly distributed between 0
and 2$\pi$ and the Fourier coefficients $a_{n}$ are normally
distributed with mean 0 and variances given by
$P(\nu) \, \delta \nu /2$ with $\nu= n \, \delta \nu = n \, \nu_0$.\\
From the long simulated time-series, light curve segments were
extracted with the same durations as the periods of continuous data
taking and with the same gaps between them. The simulated fluxes
were further smeared according to measurement errors, according to the
observing conditions (essentially zenith angle and background rate effects) in the
corresponding data set.

\subsection{Characterization of the lognormal process by the excess r.m.s.--flux relation}\label{exc_sim}

For a fixed PSD, characterized by a set of parameters $ \left\{
\alpha, K \right\}$, 500 light curves were simulated, reproducing
the observing conditions of the flaring period (MJD\,53944--53947), 
according to the procedure explained in
Section~\ref{simsec}.

\begin{figure}[tbh]
\centering
\includegraphics[width=0.54 \textwidth]{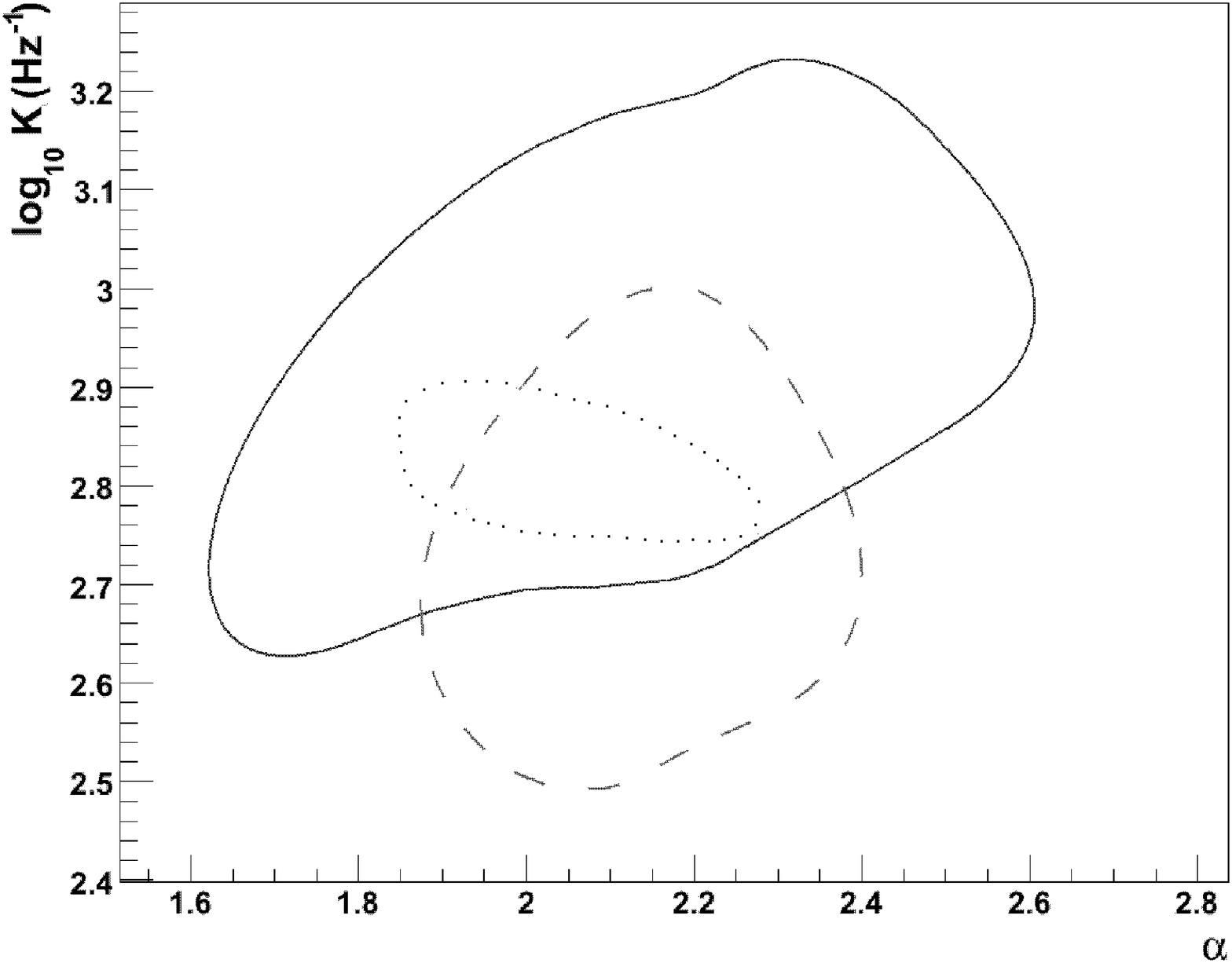} 
\caption{95\% confidence domains for $\alpha$ and $K$ at $\nu_{\rm
ref}=10^{-4}\,\mathrm{Hz}$ obtained by a maximum-likelihood method based on
the $\sigma_\mathrm{xs}$-flux correlation from 500 simulated light curves. 
The dashed contour refers to light curve segments of 20 minutes
duration, sampled every minute. The solid contour refers to light
curve segments of 80 minutes duration, sampled every 4 minutes.
The dotted contour refers to the method based on the structure function,
as explained in Section~\ref{str_fun}.} \label{cont_excess}.
\end{figure}

For each set of simulated light curves, segments of 20 minutes
duration sampled every minute (and alternatively segments of 80
minutes duration sampled every 4 minutes) were extracted and, for
each of them, the excess r.m.s. $\sigma_{xs}$ and the mean flux
$\bar{\Phi}$ were calculated as explained in Section~\ref{rmssect}.
For a large range of values of $\alpha$ and $K$, simulated light
curves reproduce well the high level of correlation found in the
measured light curves. On the other hand, the fractional
variability $F_\mathrm{var}$ and $\bar{\Phi}$ are essentially uncorrelated
and will be used in the following. A likelihood function of $\alpha$
and $K$ was obtained by comparing the simulated distributions of
$F_\mathrm{var}$ and $\bar{\Phi}$ to the experimental ones. An additional
observable which is sensitive to $\alpha$ and $K$ is the fraction of those
light curve segments for which $F_\mathrm{var}$ cannot be calculated
because large measurement errors lead to a negative value for the
excess variance. The comparison between the measured value of
this fraction and those obtained from simulations is also taken into
account in the likelihood function. The $95\%$ confidence contours
for the two parameters $\alpha$ and $K$ obtained from the maximum
likelihood method are shown in Fig.~\ref{cont_excess} for both kinds
of light curve segments. The two selected domains in the $\left\{
\alpha, K \right\}$ plane have a large overlap which restricts the
values of $\alpha$ to the interval (1.9, 2.4).

\subsection{Characterization of the lognormal process by the structure function analysis}\label{str_fun}

Another method for determining $\alpha$ and $K$ is based on
Kolmogorov structure functions (SF). For a signal $X(t)$, measured
at $N$ pairs of times separated by a delay $\tau$, $\left\{ t_i, t_i
+ \tau \right\} \, (i=1,...,N)$, the first-order structure function
is defined as (\cite{simonetti}):
\begin{eqnarray}
SF(\tau)=\frac{1}{N}\sum_{i=1}^{N}[X(t_i)-X(t_i+\tau)]^2  \\[0.mm]\nonumber
=\frac{1}{N}\sum_{i=1}^{N}[\ln \Phi(t_i)-\ln \Phi(t_i+\tau)]^2\label{sf_firstorder}
\end{eqnarray}
In the present analysis, $X(t)$ represents the variable whose PSD is
being estimated, namely $\ln \Phi$. The structure function is a powerful tool 
for studying aperiodic signals (\cite{rutman}, \cite{simonetti}), such as
the luminosity of blazars at various wavelengths. Compared to the
direct Fourier analysis, the SF has the advantage of being 
less affected by ``windowing effects'' caused by large gaps between short observation periods in VHE
observations. The first-order structure
function is adapted to those PSDs whose variability spectral index
is less than 3 \cite{rutman}, which is the case here, according to the results of
the preceding section.

\begin{figure}[tbh]
\centering
\includegraphics[width=0.535 \textwidth]{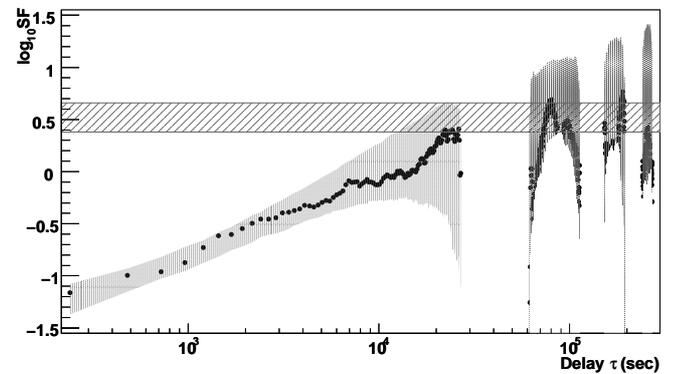}
\caption{First order structure function SF for the observations carried out in the
period MJD\,53944--53947 (circles). The shaded area corresponds to the 68\%
confidence limits obtained from simulations for the lognormal
stationary process characterized by $\alpha=2$ and $\log_{10}(K/{\rm
Hz}^{-1})=2.8$. The superimposed horizontal band represents the allowed range for the asymptotic value of the SF as obtained in Section~\ref{characteristic_time}.}
 \label{sf_4nights}
\end{figure}

Fig.~\ref{sf_4nights} shows the first-order SF estimated for the
flaring period (circles) for $\tau < 60$~hr. At fixed $\tau$,
the distribution of $\mathrm{SF}(\tau)$ expected for a given set of
parameters $\left\{\alpha, K \right\}$ is obtained from 500
simulated light curves. As an example, taking $\alpha=2$ and
$\log_{10} (K/{\rm Hz}^{-1})=2.8$, values of $\mathrm{SF}(\tau)$ are found to
lie at 68\% confidence level within the shaded region in Fig.~\ref{sf_4nights}.

In the case of a power law PSD with index $\alpha$, the SF averaged
over an ensemble of light curves is expected show a variation as
$\tau^{\alpha-1}$ (\cite{kataoka}). However, this property does not
take into account the effect of measurement errors, nor of the limited
sensitivity of Cherenkov telescopes at lower fluxes. For the present
study, it was preferable to use the distributions of $\mathrm{SF}(\tau)$
obtained from realistic simulations including all experimental
effects. Using such distributions expected for a given set of
parameters $\left\{\alpha, K \right\}$, a likelihood function can be
obtained from the experimental SF and further maximized with respect to
these two parameters. Furthermore, the likelihood fit was restricted
to values of $\tau$ lower than $10^4\,\mathrm{s}$, for which the expected
fluctuations are not too large and are well-controlled. The 95\%
confidence region in the $\left\{\alpha, K \right\}$ plane thus
obtained is indicated by the dotted line in Fig.~\ref{cont_excess}.
It is in very good agreement with those based on the excess
r.m.s.--flux correlation and give the best values for $\alpha$ and
$K$:
\begin{equation}
\alpha = 2.06 \pm 0.21 \: \: \: \mbox{and}  \: \: \:
\log_{10}(K/{\rm Hz}^{-1}) = 2.82 \pm 0.08 \label{eq:alpk}
\end{equation}
The variability index $\alpha$ at VHE energies is found to be
remarkably close to those measured in the X-ray domain on
\pks, Mrk\,421, and Mrk\,501 (\cite{kataoka}).

\subsection{Characterization of the lognormal process by doubling times}\label{dbl}

Simulations were also used to investigate if the estimator $T_2$ can
be used to constrain the values of $\alpha$ and $K$. However, for
$\alpha$ less than 2, no significant constraints on those parameters
are obtained from the values of $T_2$. For higher values of
$\alpha$, doubling times only provide loose confidence intervals on
$K$ which are compatible with the values reported above. This can be
seen in Table~\ref{tab_t2_stoch}, showing the 68\% confidence
intervals predicted for $T_2$ and $\tilde{T_2}$ for a lognormal
process with $\alpha$=2 and $\log_{10}(K/(\rm Hz^{-1}))=2.8$, as
obtained from simulation. Therefore, the variability of \pks\ 
during the flaring period can be consistently described by the
lognormal random process whose PSD is characterized by the
parameters given by Eq.~\ref{eq:alpk}.
\begin{table}[!]
\caption{Confidence interval at 68\% c.l. for $T_2$ and
$\tilde{T_2}$ predicted by simulations for $\alpha$=2 and
$\log_{10}(K/({\rm Hz}^{-1}))=2.8$ for the two high-intensity nights
MJD\,53944 and MJD\,53946, with two different sampling intervals (1
and 2 minutes).} \label{tab_t2_stoch} \centering
\begin{tabular}{c c c c}
\hline
&&&\\\vspace{1.9ex}
MJD & Bin size  & $T_2$[min] & $\tilde{T_2}$[min]  \\
\hline
53944 & 1 min & 0.93-1.85  & 1.60-2.60            \\
53944 & 2 min & 3.01-4.28  & 4.52-6.40            \\
\hline
53946 & 1 min & 1.8-2.3 &  1.96-2.41         \\
53946 & 2 min & 5.3-9.1  & 6.6-12.1           \\
\hline
\end{tabular}
\end{table}
\section{Limits on characteristic time of \pks}  \label{characteristic_time}

In Section~\ref{t2sect} the shortest variability time 
scale of \pks\ using estimators like doubling times have been estimated. 
This corresponds to exploring the high frequency behaviour of the PSD. 
In this section the lower ($<\,10^{-4}\,\mathrm{Hz}$) 
frequency part of the PSD will be considered, aiming to set a 
limit on the timescale above which the PSD, characterized 
in Section~\ref{log_normal_process}, starts to steepen to $\alpha\,>\,2$. 
A break in the PSD is expected to avoid infrared divergences and the 
time at which this break occurs can be considered as a characteristic time, from 
which physical mechanisms occurring in AGN could be inferred.

\begin{table}
\caption{Variability estimators (definitions in Section~\ref{fvar}) relative to $\ln \Phi~$ both
for the ``quiescent'' and flaring regime, as defined in Section~\ref{QuiescentStateDistrib}. 
Experimental variance (line 1), error contribution to the variance (line 2), and excess
variance (line 3). The latter is directly comparable to
$F_{var}^2$.}\label{dbg_variances} \centering
\begin{tabular}{c c c}
\hline
 & ``Quiescent'' regime & Flaring regime \\ \hline
 $\sigma^2_{\rm exp} \: \mbox{of} \: \ln \Phi$ & 0.304 $\pm$ 0.040 & 1.78
 $\pm$
 0.27 \\
$\sigma^2_{\rm err} \: \mbox{of} \: \ln \Phi$ & 0.169 $\pm$ 0.053 &
0.022 $\pm$
 0.005 \\
$\sigma^2_{\rm xs} \: \mbox{of} \: \ln \Phi$ & 0.135 $\pm$ 0.067 &
1.758 $\pm$ 0.273
 \\
\hline
\end{tabular}
\end{table}

Clearly the description of the source variability during the flaring period 
by a stationary lognormal random process
is in good agreement with the flux distributions shown in Fig.~\ref{doubleGaussian_fit}.
Considering the second Gaussian fit in the right panel of Fig.~\ref{doubleGaussian_fit}, 
the excess variance in
the flaring regime reported in Table~\ref{dbg_variances},
although affected by a large error, is an
estimator of the intrinsic variance of the stationary process. It
has been demonstrated that $2\,\sigma^2_{\rm xs}$ represents the
asymptotic value of the first-order structure function for large
values of the delay $\tau$ (\cite{simonetti}). On the other hand, as
already mentioned, a PSD proportional to $\nu^{-\alpha}$ with
$\alpha \approx 2$ cannot be extrapolated to arbitrary low
frequencies; equivalently, the average structure function cannot
rise as $\tau^{\alpha-1}$ for arbitrarily long times. Therefore, by
setting a 95\% confidence interval on $\log_{10} \mathrm{SF}_{\rm asympt} =
\log_{10}(2 \, \sigma^2_{\rm xs})$ of $\left[0.38, 0.66 \right]$
from Table \ref{dbg_variances}, it is possible to evaluate a confidence
interval on a timescale above which the average value of the
structure function cannot be described by a power law. Taking
account of the uncertainties on $\alpha$ and $K$ given by Eq.~\ref{eq:alpk}, leads to the 95\% confidence interval for this
characteristic time $T_{\rm char}$ of the blazar in the flaring
regime:
\[ 3 \: \mbox{hours} < T_{\rm char} < 20 \: \mbox{hours} \]
This is compatible with the behaviour of the experimental structure
function at times $\tau > 10^4$~s (Fig.~\ref{sf_4nights}), although
the large fluctuations expected in this region do not allow a more
accurate estimation. In the X-ray domain, characteristic times of
the order of one day or less have been found for several blazars
including \pks\ (\cite{kataoka}). The results presented here
suggest a strong similarity between the PSDs for X-rays and VHE
$\gamma$-rays during flaring periods.

\section{Discussion and conclusions}\label{conclusion}

This data set, which exhibits unique features and results, is the outcome of a long-term
monitoring program and dedicated, dense, observations. One of the main results here is the
evidence for a VHE $\gamma$-ray quiescent-state emission, where the variations in the flux 
are found to have a lognormal distribution. The existence of such a state was
postulated by \cite{ste96} in order to explain the extragalactic $\gamma$-ray background at
0.03--$100\,{\rm GeV}$ detected by EGRET (\cite{fic96,sre98}) as coming from quiescent-state
unresolved blazars. Such a background has not yet been detected in the VHE range, as it is technically difficult 
with the atmospheric Cherenkov technique to find an isotropic extragalactic emission and even more to 
distinguish it from the cosmic-ray electron flux (\cite{egb08}).
In addition, the EBL attenuation limits the distance from which $\sim$TeV
$\gamma$-rays can propagate to $\sim 1\,{\rm Gpc}$ (\cite{HESSEBL}). As pointed out by
\cite{che00}, emission mechanisms might be simpler to understand during quiescent states in
blazars, and they are also the most likely state to be found observationally. In the X-ray
band, the existence of a steady underlying emission has also been invoked for two other VHE
emitting blazars (Mrk~421, \cite{fos00}, and 1ES~1959+650, \cite{gie02}). Being able to
separate, and detect, flaring and nonflaring states in VHE $\gamma$-rays is hence important for such studies.

The observation of the spectacular outbursts of \pks\ in July 2006 represents one the most
extreme examples of AGN variability in the TeV domain, and allows spectral and timing
properties to be probed over two orders of magnitude in flux.

Whereas for the flaring states with fluxes above a few $10^{-10}\,\percmsqrs$ 
a clear hardening of the spectrum with increasing flux is observed, familiar also from the blazars Mrk 421 
and Mrk 501, for the quiescent state in contrast an indication of a softening is noted. 
If confirmed, this is a new and intruiging observation in the VHE regime of blazars.
The blazar PKS~0208$-$512 (of the FSRQ class) also shows such initial softening and subsequent
hardening with flux in the MeV range, but no general trend could be found for $\gamma$-ray
blazars (\cite{nan07}). In the framework of synchrotron self-Compton scenarios, VHE spectral
softening with increasing flux can be associated with, for example, an increase in magnetic field
intensity, emission region size, or the power law index of the underlying electron
distribution, keeping all other parameters constant. A spectral hardening can equally be
obtained by increasing the maximal Lorentz factor of the electron distribution or the Doppler
factor (see e.g. Fig.~11.7 in \cite{kat99}). A better understanding of the mechanisms at play
would require multi-wavelength observations of similar time span and sampling density as the
data set presented here.

It is shown that the variability time scale $t_{\rm var}$ of a few minutes are only upper
limits for the intrinsic lowest characteristic time scale. Doppler factors of $\delta \geq
100$ of the emission region are derived by \cite{HESS2155_BigFlare}~(2007a) using the
$\sim$~$10^9~M_\odot$ black hole (BH) Schwarzschild radius light crossing time as a limit, while
\cite{begelman} argue that such fast time scales cannot be linked to the size of the BH and
must occur in regions of smaller scales, such as ``needles'' of matter moving faster than
average within a larger jet (\cite{ghi08}), small components in the jet dominating at TeV
energies (\cite{kat08}), or jet ``stratification'' (\cite{bou08}). \cite{levinson} attributes
the variability to dissipation in the jet coming from radiative deceleration of shells with
high Lorentz factors.

The flaring period allowed the study of light curves in separated energy bands and the
derivation of a power law dependence of $F_{\rm var}$ with the energy ($F_{\rm
  var}\propto E^{\sim 0.2}$). This dependence is comparable to that reported in
\cite{berrie}~(2007), \cite{integr_mrk}~(2008), \cite{xmm}~(2002), where $F_{\rm var}(E)\propto E^{\sim0.2}$ between the optical
and X-ray energy bands was found for Mrk\,421 and \pks, respectively. An increase with the
energy of the flux variability has been found for Mrk\,501 (\cite{magic_mrk501}~2007) in VHE
$\gamma$-rays on timescales comparable to those observed here.

 The flaring period showed for the first time that the intrinsic variability of \pks\ increases
 with the flux, which can itself be described by a lognormal process, indicating that the
 aperiodic variability of \pks\ could be produced by a multiplicative process. The flux in the
 ``quiescent regime'', which is on average 50 times lower than in the flaring period and
 has a 3 times lower $F_{\rm var}$, also follows a lognormal distribution, suggesting
 similarities between these two regimes.

It has been possible to characterize a power spectral density of the flaring period in the
frequency range $10^{-4}$--$10^{-2}\,{\rm Hz}$, resulting in a power law of index
$\alpha=2.06\pm0.21$ valid for frequencies down to $\sim 1/{\rm day}$. The description of the
rapid variability of a TeV blazar as a random stationary process must be taken into account by
time-dependent blazar models. For \pks~the evidence of this log-normality has been found very recently in X-rays 
(\cite{pks_xrays}) and  as previously mentioned, X-ray binaries and Seyfert galaxies also
show lognormal variability, which is thought to originate from the accretion disk
(\cite{mchardy04,lyubarskii,arevalo}), suggesting a connection between the disk and the
jet. 
This variability behaviour should therefore be searched for in existing blazar light
curves, independently of the observed wavelength.

\section*{Acknowledgements}
The support of the Namibian authorities and of the University of Namibia
in facilitating the construction and operation of H.E.S.S. is gratefully
acknowledged, as is the support by the German Ministry for Education and
Research (BMBF), the Max Planck Society, the French Ministry for Research,
the CNRS-IN2P3 and the Astroparticle Interdisciplinary Programme of the
CNRS, the U.K. Science and Technology Facilities Council (STFC),
the IPNP of the Charles University, the Polish Ministry of Science and
Higher Education, the South African Department of
Science and Technology and National Research Foundation, and by the
University of Namibia. We appreciate the excellent work of the technical
support staff in Berlin, Durham, Hamburg, Heidelberg, Palaiseau, Paris,
Saclay, and in Namibia in the construction and operation of the
equipment.

\appendix

\section{Observations summary}
\label{JournalNights}
The journal of observations for the 2005-2007 is presented in Table~\ref{JournalNightsTab}.

\begin{table*}[!]
\begin{center}
\caption{
\label{JournalNightsTab}
Summary of the 2005 to 2007 observations, where for each night {\it MJD} is the Modified Julian Date,
{\it $\theta_{z}$} the mean observation zenith angle (degrees), {\it T} the total observation live-time (hours), 
{\it $n_\mathrm{on}$} the number of on-source events, {\it $n_\mathrm{off}$} the number of off-source events
(from a region five times larger than for the on-source events).
The final three columns are the corresponding excess, significance (given in units of standard deviations), and the significance per square root of the live-time.
}
\begin{tabular*}{0.52\textwidth}{*{8}{c}}
   \hline
MJD   & $\theta_{z}$  & $T$   &   $n_\mathrm{on}$ &  $n_\mathrm{off}$ &   Excess&  $\sigma$ & $\sigma/\sqrt{T}$\\ 
   \hline
53618 & 16.9 & 0.87 &   860 & 3,202 &   219.6 &   7.5 &   8.0 \\
53637 & 16.6 & 1.09 &   788 & 2,673 &   253.4 &   9.2 &   8.8 \\
53638 & 14.7 & 2.18 & 1,694 & 6,029 &   488.2 &  12.0 &   8.1 \\
53665 & 22.6 & 0.87 &   618 & 2,487 &   120.6 &   4.7 &   5.1 \\
53666 & 26.3 & 1.31 &   857 & 3,406 &   175.8 &   5.9 &   5.1 \\
53668 & 23.3 & 1.30 &   926 & 3,793 &   167.4 &   5.3 &   4.7 \\
53669 & 19.6 & 0.86 & 1,027 & 2,939 &   439.2 &  14.7 &  15.8 \\
53705 & 55.6 & 0.88 &   512 & 2,542 &     3.6 &   0.1 &   0.2 \\
53916 & 13.7 & 0.88 &   993 & 3,317 &   329.6 &  10.7 &  11.5 \\
53917 & 11.8 & 0.88 &   933 & 3,163 &   300.4 &  10.1 &  10.7 \\
53918 & 10.2 & 1.32 & 1,491 & 4,596 &   571.8 &  15.5 &  13.5 \\
53919 & 10.9 & 1.32 & 1,477 & 4,638 &   549.4 &  14.9 &  13.0 \\
53941 & 14.4 & 1.31 & 2,445 & 4,844 & 1,476.2 &  35.0 &  30.6 \\
53942 & 13.7 & 1.76 & 2,453 & 5,766 & 1,299.8 &  29.5 &  22.3 \\
53943 &  9.8 & 1.33 & 1,142 & 3,627 &   416.6 &  12.8 &  11.1 \\
53944 & 13.2 & 1.33 &12,762 & 3,563 &12,049.4 & 172.9 & 149.7 \\
53945 & 23.9 & 5.23 & 8,037 &16,352 &  4766.6 &  62.0 &  27.1 \\
53946 & 27.7 & 6.61 &35,874 &19,881 &31,897.8 & 251.3 &  97.7 \\
53947 & 25.1 & 5.89 &17,158 &17,006 &13,756.8 & 142.6 &  58.8 \\
53948 & 27.7 & 2.75 & 5,366 & 7,957 & 3,774.6 &  64.6 &  38.9 \\
53950 & 26.6 & 3.51 & 5,108 &11,955 & 2,717.0 &  42.8 &  22.9 \\
53951 & 28.4 & 2.51 & 3,275 & 8,421 & 1,590.8 &  30.6 &  19.3 \\
53952 & 35.8 & 1.76 & 1,786 & 5,395 &   707.0 &  17.7 &  13.3 \\
53953 & 44.1 & 0.89 &   670 & 2,285 &   213.0 &   8.4 &   8.9 \\
53962 & 27.6 & 0.89 &   534 & 2,088 &   116.4 &   4.9 &   5.3 \\
53963 & 19.4 & 1.75 & 1,613 & 6,145 &   384.0 &   9.5 &   7.1 \\
53964 & 10.3 & 1.49 & 1,057 & 4,146 &   227.8 &   6.9 &   5.6 \\
53965 & 15.7 & 1.57 & 1,584 & 5,662 &   451.6 &  11.4 &   9.1 \\
53966 & 18.6 & 0.88 &   719 & 2,844 &   150.2 &   5.5 &   5.8 \\
53967 & 24.3 & 0.86 &   481 & 1,801 &   120.8 &   5.5 &   5.9 \\
53968 & 19.1 & 0.86 &   479 & 1,974 &    84.2 &   3.7 &   4.0 \\
53969 & 21.2 & 1.29 & 1,738 & 4,368 &   864.4 &  23.0 &  20.2 \\
53970 & 20.9 & 0.88 &   690 & 2,759 &   138.2 &   5.1 &   5.5 \\
53971 & 19.8 & 1.32 & 1,449 & 4,313 &   586.4 &  16.3 &  14.2 \\
53972 & 16.5 & 1.30 &   683 & 2,499 &   183.2 &   7.0 &   6.1 \\
53973 & 14.5 & 1.32 & 1,157 & 4,311 &   294.8 &   8.6 &   7.5 \\
53974 & 16.2 & 1.76 & 1,504 & 5,925 &   319.0 &   8.1 &   6.1 \\
53975 & 15.8 & 0.88 &   804 & 3,059 &   192.2 &   6.7 &   7.2 \\
53976 & 13.8 & 0.89 &   727 & 2,544 &   218.2 &   8.2 &   8.7 \\
53977 & 12.0 & 0.88 &   832 & 2,745 &   283.0 &  10.1 &  10.8 \\
53978 & 13.3 & 0.89 &   687 & 2,317 &   223.6 &   8.7 &   9.3 \\
53995 & 21.4 & 1.75 & 1,712 & 5,989 &   514.2 &  12.6 &   9.5 \\
53996 & 25.3 & 1.33 &   680 & 2,834 &   113.2 &   4.2 &   3.6 \\
53997 & 26.3 & 1.32 & 1,113 & 4,382 &   236.6 &   7.0 &   6.1 \\
53998 & 26.1 & 1.32 & 1,247 & 4,082 &   430.6 &  12.6 &  11.0 \\
53999 & 21.4 & 1.32 & 1,107 & 4,262 &   254.6 &   7.5 &   6.6 \\
54264 &  8.5 & 0.13 &   109 &   449 &    19.2 &   1.8 &   5.0 \\
54265 &  8.5 & 1.05 &   920 & 3,513 &   217.4 &   7.1 &   6.9 \\
54266 &  9.3 & 1.39 & 1,129 & 4,553 &   218.4 &   6.3 &   5.4 \\
54267 &  9.2 & 1.32 & 1,095 & 4,176 &   259.8 &   7.8 &   6.7 \\
54268 & 10.2 & 0.32 &   261 & 1,040 &    53.0 &   3.2 &   5.6 \\
54269 &  9.1 & 0.89 &   908 & 2,491 &   409.8 &  14.7 &  15.6 \\
54270 & 10.8 & 0.44 &   565 & 1,266 &   311.8 &  14.9 &  22.4 \\
54271 &  7.6 & 0.36 &   308 &   983 &   111.4 &   6.6 &  11.0 \\
54294 &  7.7 & 0.44 &   447 & 1,337 &   179.6 &   9.0 &  13.5 \\
54296 &  7.2 & 0.44 &   350 & 1,289 &    92.2 &   4.9 &   7.4 \\
54297 &  9.8 & 0.44 &   344 & 1,265 &    91.0 &   4.9 &   7.4 \\
54299 &  7.6 & 0.44 &   347 & 1,232 &   100.6 &   5.5 &   8.2 \\
54300 &  7.3 & 0.44 &   343 & 1,245 &    94.0 &   5.1 &   7.6 \\
54302 &  7.7 & 0.44 &   358 & 1,234 &   111.2 &   6.0 &   9.0 \\
54304 &  7.9 & 0.88 &   680 & 2,765 &   127.0 &   4.7 &   5.0 \\
54319 &  8.9 & 0.88 &   692 & 2,650 &   162.0 &   6.1 &   6.5 \\
54320 &  8.0 & 0.89 &   553 & 2,199 &   113.2 &   4.7 &   5.0 \\
54329 & 11.7 & 0.44 &   297 & 1,258 &    45.4 &   2.5 &   3.8 \\
54332 &  7.0 & 0.16 &   100 &   391 &    21.8 &   2.1 &   5.3 \\
54375 &  9.4 & 1.27 &   811 & 3,124 &   186.2 &   6.4 &   5.7 \\
54376 &  7.9 & 0.68 &   395 & 1,605 &    74.0 &   3.6 &   4.4 \\
   \hline
\end{tabular*}
\end{center}
\end{table*}

\clearpage

\section{Spectral variability}
\label{SpecVarApp}

The numerical information associated with Fig.~\ref{SpecVar20052007_mix} is given in 
Tables~\ref{SpecVarFTabLeftGrey} (left panel, grey points), \ref{SpecVarFTabLeftBlack}
(left panel, black points) and \ref{SpecVarFTabRight} (right panel). In addition, numerical information associated 
with Fig.~\ref{FourNightsLC} is given in Table~\ref{SpecVar4nightsTab}.

\begin{table}[!h]
\begin{center}
\caption{
\label{SpecVarFTabLeftGrey}
Integral flux ($10^{-11}\,\percmsqrs$) in the 0.2--1\,TeV energy range versus photon 
index corresponding to grey points in the left panel of Fig.~\ref{SpecVar20052007_mix}. 
Errors are statistical. See Section~\ref{SpecVarAll} for more details.} 
\begin{tabular}{*{2}{c}}
   \hline
$\Phi$ & Index $\Gamma$ \\ 
   \hline
$2.36\pm0.13$ & $3.345\pm0.20$ \\
$3.92\pm0.17$ & $3.64\pm0.16$ \\
$5.33\pm0.22$ & $3.46\pm0.13$ \\
$8.29\pm0.30$ & $3.64\pm0.10$ \\
$13.82\pm0.82$ & $3.82\pm0.17$ \\
   \hline
\end{tabular}
\end{center}
\end{table}

\begin{table}[!h]
\begin{center}
\caption{
\label{SpecVarFTabLeftBlack}
Integral flux ($10^{-11}\,\percmsqrs$) in the 0.2--1\,TeV energy range versus photon 
index corresponding to black points in the left panel of Fig.~\ref{SpecVar20052007_mix}. 
Errors are statistical. See Section~\ref{SpecVarAll} for more details.} 
\begin{tabular}{*{2}{c}}
   \hline
$\Phi$ & Index $\Gamma$ \\ 
   \hline
$8.4\pm0.3$ & $3.74\pm 0.11$ \\
$16.9\pm0.5$ & $3.82\pm 0.10$ \\
$24.5\pm0.7$ & $3.78\pm 0.08$ \\
$37.4\pm1.0$ & $3.77\pm 0.08$ \\
$39.7\pm1.1$ & $3.76\pm 0.08$ \\
$46.4\pm1.1$ & $3.66\pm 0.08$ \\
$53.5\pm1.3$ & $3.57\pm 0.07$ \\
$78.6\pm1.9$ & $3.44\pm 0.06$ \\
$91.8\pm1.9$ & $3.33\pm 0.06$ \\
$101.6\pm2.8$ & $3.30\pm 0.07$ \\
$111.7\pm3.0$ & $3.33\pm 0.07$ \\
$154.1\pm3.5$ & $3.28\pm 0.06$ \\
$173.1\pm3.8$ & $3.16\pm 0.06$ \\
$198.5\pm3.8$ & $3.28\pm 0.05$ \\
$210.9\pm3.9$ & $3.14\pm 0.05$ \\
   \hline
\end{tabular}
\end{center}
\end{table}

\begin{table}[!h]
\begin{center}
\caption{
\label{SpecVarFTabRight}
Integral flux ($10^{-11}\,\percmsqrs$) in the 0.2--1\,TeV energy range versus photon 
index corresponding to the right panel of Fig.~\ref{SpecVar20052007_mix}. 
Errors are statistical. See Section~\ref{SpecVarAll} for more details.} 
\begin{tabular}{*{2}{c}}
   \hline
$\Phi$ & Index $\Gamma$ \\ 
   \hline
$4.2\pm0.1$ & $3.52\pm 0.07$ \\
$17.3\pm0.5$ & $3.89\pm 0.09$ \\
$38.6\pm1.1$ & $3.80\pm 0.08$ \\
$43.6\pm1.1$ & $3.60\pm 0.07$ \\
$51.5\pm1.3$ & $3.53\pm 0.07$ \\
$67.2\pm1.9$ & $3.64\pm 0.07$ \\
$86.1\pm1.9$ & $3.38\pm 0.06$ \\
$97.0\pm1.9$ & $3.30\pm 0.05$ \\
$111.7\pm3.0$ & $3.33\pm 0.07$ \\
$154.1\pm3.5$ & $3.28\pm 0.06$ \\
$173.1\pm3.8$ & $3.16\pm 0.06$ \\
$198.5\pm3.8$ & $3.28\pm 0.05$ \\
$210.9\pm3.9$ & $3.14\pm 0.04$ \\
   \hline
\end{tabular}
\end{center}
\end{table}

\begin{table}[!h]
\begin{center}
\caption{
\label{SpecVar4nightsTab}
MJD, integral flux ($10^{-11}\,\percmsqrs$) in the 0.2--1\,TeV energy range, and photon index corresponding
to the entries of Fig~\ref{FourNightsLC}. Only points associated with an energy threshold lower than $200\,{\rm GeV}$
are considered. Errors are statistical. See Section~\ref{SpecFlaringNights} for more details.} 
\begin{tabular}{*{3}{c}}
\hline
MJD & $\Phi$ & Index $\Gamma$ \\ 
   \hline
$53944.02742\pm0.00277$ & $188.6\pm30.6$ & $3.22\pm0.09$ \\
$53944.03298\pm0.00277$ & $184.1\pm30.9$ & $3.28\pm0.09$ \\
$53944.03854\pm0.00277$ & $191.7\pm32.6$ & $3.45\pm0.09$ \\
$53944.04409\pm0.00277$ & $252.4\pm40.5$ & $3.19\pm0.09$ \\
$53944.04965\pm0.00277$ & $237.5\pm33.8$ & $3.16\pm0.08$ \\
$53944.05520\pm0.00277$ & $212.8\pm30.8$ & $3.04\pm0.08$ \\
$53944.06076\pm0.00277$ & $190.9\pm30.0$ & $3.09\pm0.09$ \\
$53944.06909\pm0.00555$ & $99.5\pm17.5$ & $3.18\pm0.10$ \\
$53944.98298\pm0.05277$ & $9.2\pm3.1$ & $3.89\pm0.18$ \\
$53945.04965\pm0.01388$ & $34.2\pm8.3$ & $3.83\pm0.13$ \\
$53945.07604\pm0.01250$ & $42.6\pm10.2$ & $3.92\pm0.13$ \\
$53945.93020\pm0.00277$ & $206.9\pm34.9$ & $3.2\pm0.10$ \\
$53945.93715\pm0.00416$ & $190.9\pm28.6$ & $3.15\pm0.09$ \\
$53945.94409\pm0.00277$ & $171.0\pm30.5$ & $3.23\pm0.10$ \\
$53945.94965\pm0.00277$ & $161.2\pm28.2$ & $3.06\pm0.10$ \\
$53945.95659\pm0.00416$ & $173.9\pm28.9$ & $3.35\pm0.09$ \\
$53945.96354\pm0.00277$ & $179.1\pm29.4$ & $3.11\pm0.10$ \\
$53945.97048\pm0.00416$ & $127.8\pm22.6$ & $3.36\pm0.10$ \\
$53945.98159\pm0.00694$ & $91.9\pm16.5$ & $3.42\pm0.10$ \\
$53945.99687\pm0.00833$ & $101.6\pm16.9$ & $3.12\pm0.10$ \\
$53946.00937\pm0.00416$ & $104.4\pm19.7$ & $3.28\pm0.11$ \\
$53946.01909\pm0.00555$ & $92.2\pm17.1$ & $3.37\pm0.10$ \\
$53946.03020\pm0.00555$ & $79.5\pm15.9$ & $3.42\pm0.11$ \\
$53946.03992\pm0.00416$ & $105.6\pm19.3$ & $3.26\pm0.10$ \\
$53946.04965\pm0.00555$ & $114.7\pm20.9$ & $3.45\pm0.10$ \\
$53946.05937\pm0.00416$ & $110.2\pm19.6$ & $3.17\pm0.10$ \\
$53946.06909\pm0.00555$ & $97.2\pm19.9$ & $3.54\pm0.11$ \\
$53946.08298\pm0.00833$ & $75.1\pm12.8$ & $3.24\pm0.10$ \\
$53946.09826\pm0.00694$ & $75.1\pm15.5$ & $3.55\pm0.12$ \\
$53946.93437\pm0.00972$ & $58.1\pm13.5$ & $3.92\pm0.13$ \\
$53946.95381\pm0.00972$ & $50.7\pm11.2$ & $3.6\pm0.12$ \\
$53946.97604\pm0.01250$ & $39.7\pm8.7$ & $3.62\pm0.12$ \\
$53947.00242\pm0.01388$ & $34.0\pm7.5$ & $3.62\pm0.12$ \\
$53947.02742\pm0.01111$ & $40.4\pm8.8$ & $3.56\pm0.12$ \\
$53947.04965\pm0.01111$ & $43.5\pm10.3$ & $3.82\pm0.13$ \\
$53947.07742\pm0.01666$ & $45.8\pm10.1$ & $3.75\pm0.12$ \\
   \hline
\end{tabular}
\end{center}
\end{table}

\end{document}